\documentclass[twocolumn, usenatbib]{mnras}
\usepackage{savesym}
\usepackage{graphicx}
\usepackage{longtable}
\usepackage{changepage}

\expandafter\let\csname equation*\endcsname\relax
  \expandafter\let\csname endequation*\endcsname\relax 
\usepackage{subfig}
\usepackage{amsmath}
\usepackage{amssymb}
\usepackage{verbatim}
\usepackage[yyyymmdd,hhmmss]{datetime}
\usepackage{array}
\usepackage{times}
\usepackage{xcolor}

\DeclareRobustCommand{\DE}[3]{#2}
\let\DEthebibliography\thebibliography
\def\thebibliography{\DeclareRobustCommand{\DE}[3]{##3}\DEthebibliography}

\newcommand{\eroszerofour}{eRASSt~J045650.3-203750}

\title[Disks in partial tidal disruption events]{ Accretion disks in (repeating) partial tidal disruption events: 
rapid state transitions, UV plateaus and flares from disk-remnant collisions  }
\author [Andrew Mummery]{Andrew Mummery$^{1, 2}$\thanks{E-mail:
amummery@ias.edu} \\
$^1$Rudolf Peierls Centre for Theoretical Physics, Beecroft Building, Parks Road, Oxford OX1 3PU, UK\\
$^2$School of Natural Sciences, Institute for Advanced Study, 1 Einstein Drive, Princeton, NJ 08540, USA
}

\date{}

\begin{document}

\maketitle

\begin{abstract}
    Tidal disruption events which repeat on timescales of months-to-years represent an unambiguous signature of a partial disruption, with the surviving stellar remnant returning to pericentre to be repeatedly stripped by tidal forces. These systems therefore offer the best laboratories to study the differences between partial and full disruptions. One noteworthy observational difference between the two systems is that all known X-ray bright repeating TDEs show rapid transitions between thermal, non-thermal and completely dim states on timescales much shorter than full (non-repeating) TDEs. We argue this can be simply understood as being due to the reduction in fuel supply available to the disk, and that these systems provide evidence that all tidal disruption events undergo a state transition at Eddington ratios $f_{\rm edd} = L_{\rm bol}/L_{\rm edd} \sim 0.01$, similar to X-ray binaries. {As part of this calculation we derive a general expression for the time taken for a TDE disk to fall to a given Eddington fraction, which will be of use to both full and partial TDE science.} Perhaps surprisingly, the late-time optical/UV plateau luminosity observed from these systems is largely unaffected by this reduction in fuel supply, provided the outer disk remains in a thermal state for long enough for this emission to be detected.  We then show that collisions between the returning stellar remnant and the  disk formed from the last passage will produce potentially observable X-ray flares ($L_{\rm flare} \simeq 10^{42}$ erg/s), but that they are likely to be very difficult to detect as they are generally short-lived ($t_{\rm flare} \simeq 0.1-1$ hr). 
\end{abstract}

\noindent

\section{Introduction}
The tidal disruption of a star by a galactic center supermassive black hole is an inevitable consequence of the high number density of stars surrounding these black holes. If a star is scattered onto a near-radial orbit which enters its own tidal radius then it will be completely destroyed by the tidal gravitational force of the black hole. If, however, the stars orbit takes it close, but still outside of, the tidal radius then a small fraction of the stellar material may be stripped from the star (and produce observable emission), with the surviving remnant potentially returning to be disrupted again on timescales of months-to-years {(although of course the timescale for returning may well be significantly longer than this, see e.g.,  \citealt{Cufari22} for discussion)}. {Tidal disruption events which are observed to repeat on short timescales are the only sources which can be unambiguously associated with partial disruptions, a result of the observed repeat timescale being significantly shorter than the typical time between two different TDEs $\Delta t_{\rm TDE} \sim 10^4-10^5$ years \citep{Magorrian99}}\footnote{The immediate post-tidal-encounter evolution of the stripped stellar debris (principally the ``fallback rate'') differs between partial and full disruptions \citep{Coughlin19}, with partial disruptions having a steeper decay in mass fallback at large times $\dot M_{\rm fb} \sim t^{-a}$, with $a=9/4$ for partial disruptions and $a=5/3$ for full disruptions. However this only describes the time evolution of the stripped mass returning to pericentre, which of course does not necessarily describe the evolution of the luminosity in any observationally relevant band (as some physical process must be invoked beyond mass returning to pericentre to produce detectable emission). The fact that the observed optical flares from tidal disruption events show a very broad range of inferred decay indices (which far exceed the theoretical differences in the fallback rate) prevents partial disruptions from being identified in this manner. See Appendix \ref{app} for more details.}. 

A small but interesting observed population of repeating partial disruptions is now taking shape, with X-ray bright sources presented in \cite{Wevers21} (the source AT2018fyk), \cite{Liu23} (the source \eroszerofour) and \cite{Saxton25} (the source XMMSL2~J1404-2511). Repeating optical tidal disruption events have also been discovered \citep{Arcavi22TNS, Somalwar23}. One interesting observational fact about these sources is that all suspected/confirmed repeating partial TDEs \citep{Wevers23fyk,Liu23, Saxton25} which are initially X-ray bright undergo state transitions, with initially thermal disk emission becoming non-thermal (i.e., ``hard''), on timescales of $\sim 100$'s of days. These sources then show large amplitude dimming episodes after another $\sim $ few hundred days into their evolution. This transition appears to be (i) deep, and (ii) rapid (or too rapid to be a complete draining of the disk), suggesting something intrinsic is happening to the disk itself. 

Such state transitions are also observed to occur routinely in another black hole disk system, namely X-ray binaries \citep[e.g.,][]{Fender04}. These transitions are typically associated with the luminosity of the system passing through universal fractions of the Eddington luminosity, with the hard-state transition usually associated with a luminosity $L_{\rm bol}/L_{\rm edd} \sim f_{\rm tr} \sim 0.01$. The transition to the quiescent state then follows at a yet lower Eddington ratio, and is typically similar to the rapid transition seen in these repeating partial TDEs. An obvious question to ask is why (repeating) partial tidal disruption events routinely undergo these transitions, while some  disruptions are observed to remain in the soft state for over 1000 days \citep[e.g., ASASSN-14li,][]{Bright18, GuoloMum24}. It is the purpose of this paper to answer this question. 

We shall argue that both of these different timescales (full and partial) are a natural prediction of time-dependent disk theory, and the observed state transitions in (repeating) partial TDEs result ultimately from a reduction in fuel supply available to the disk, and that these systems provide evidence that all tidal disruption events undergo a state transition at Eddington ratios $f_{\rm edd} \sim 0.01$. The natural timescale for the same transitions to occur in full disruptions is $\sim$ decades, in keeping with current observations of the majority of the TDE population {(although this timescale can be much shorter for high mass black holes $M_\bullet \sim 10^8M_\odot$)}.  We then go on to show that, perhaps surprisingly, the late-time UV ``plateau'' luminosity (a near time-independent phase of TDE optical/UV evolution dominated by disk emission) roughly $\sim$ years post disruption shows only a weak dependence on this reduction in fuel supply, with full disruptions brightest (as can be understood analytically). This is important because it implies that scaling relationships between the UV plateau luminosity and black hole mass \citep{Mummery_et_al_2024, Mummery26flare} which are calibrated on full disruptions still hold for partial disruptions, removing a potential systematic uncertainty. 

Finally, another natural question to ask is whether there may be any observational consequences of the interaction between the accretion flow formed from the last stellar-passing and the returning stellar remnant. We consider the possible observational consequences of the evolution of shock-heated gas produced by a collision between the returning stellar remnant and the disk. The collisions produce reasonably bright flares $L_{\rm flare} \sim 10^{42}$ erg/s, but they are typically very short lived $t_{\rm flare} \sim 0.1-1$ hours. We do not expect these flares to be detected, except in extremely fortuitous circumstances. 

The layout of this paper is as follows. In section \ref{theory1} we derive the typical timescales for partial and full disruptions to undergo state transitions (more precisely, to reach a fixed Eddington fraction), before confirming these results with numerical experiments in section \ref{num}. In section \ref{theory2} we consider the possible observable consequences of collisions between the existing disk and the stellar remnant, upon the return of the remnant.  We conclude in section \ref{conc}. A discussion of fallback rates and optical TDE emission is presented in Appendix \ref{app}. 

\section{Properties of partial-TDE disks}\label{theory1}
\subsection{Stripped mass}
A partial TDE occurs when the pericentre of the incoming stars orbit $(r_p)$ is greater than, but ultimately close to, the ``canonical'' tidal radius $(r_T)$ of the star. Defining the usual impact parameter $\beta$
\begin{equation}
    r_p = {r_T\over \beta} = {R_\star \over \beta} \left({M_\bullet \over M_\star}\right)^{1/3} .
\end{equation}
{A second key radial scale must be defined, which we shall denote ${\cal R}_T$. The radius ${\cal R}_T$ represents the orbital radius at which the incoming star is completely disrupted. For almost all stars this distance is smaller than the ``canonical'' tidal radius of the star $r_T$ defined above \citep{Coughlin22b}. We follow convention and define the critical impact parameter $\beta_c$ at which a full disruption occurs, which  in our notation is equal to }
\begin{equation}
    \beta_c \equiv {r_T \over {\cal R}_T} =  {R_\star \over {\cal R}_T}   \left({M_\bullet \over M_\star}\right)^{1/3} .
\end{equation}
Partial TDEs imply, therefore,  $\beta \leq \beta_c$. {The precise value of $\beta_c$ depends on the structure of the star itself. } In the case of a full disruption ($r_{ p} \le {\cal R}_{ T})$, we shall assume that half of the stellar mass falls back as bound debris, $M_{\rm fb} = M_{\star}/2$ \citep{Rees88}. For partial disruptions ($r_{p} > {\cal R}_{ T})$, we {can gain insight by considering} the results of \cite{Ryu+20a}. \cite{Ryu+20a} ran hydrodynamic simulations of the disruption process and found that the total mass ``falling back'' to potentially form a disk satisfied 
\begin{equation}
    M_{\rm fb} = \frac{1}{2}\left(M_{\star}-M_{\rm rem}\right),
\end{equation}
where 
\begin{equation}
    \frac{M_{\rm rem}}{M_{\star}} = 1 - \left(\frac{r_{ p}}{{\cal R}_{ T}}\right)^{-3} ,
\end{equation}
is the mass of the bound remnant. Taking this result literally {would} imply that the {\it maximum} total debris mass available to accrete in a partial TDE is 
\begin{equation}
    M_{\rm fb} = {1 \over 2}\left({\beta \over \beta_c}\right)^3 M_\star, \quad\quad \beta \leq \beta_c.
\end{equation}
{We note at this point that the factor $1/2$ normalisation here implicitly assumes that the incoming star was on a parabolic orbit, something which will not be the case when the star returns for repeat disruptions (for it must be on an elliptical orbit at this stage). For an elliptical orbit more than one half of the stripped debris will be bound to the black hole, with the fraction of the debris remaining bound increasing with dropping eccentricity of the incoming orbit\footnote{I am grateful to the referee for pointing this out.}. }

Other $\beta$-dependent mass fallback profiles have been presented in other works, notably \cite{Guillochon13} and \cite{Chen24}. The more recent work \citep{Chen24} find power-law profiles in keeping with \cite{Ryu+20a}, namely $\sim \beta^{2.5}-\beta^{4}$. 
We stress that it is expected \citep[a result confirmed by simulations][]{Chen24} that this surviving mass fraction is expected to be only dependent on the ratio $\beta/\beta_c$ (and therefore depends on internal stellar structure via $\beta_c$), but not otherwise on black hole mass or stellar mass.  {For the remainder of this work, to be flexible, we consider a general scaling of}
\begin{equation}
    M_{\rm fb} = {1 \over 2}\left({\beta \over \beta_c}\right)^q M_\star, \quad\quad \beta \leq \beta_c ,
\end{equation}
{but will often assume $q=3$ for the purposes of making Figures.}
\subsection{Surface density on a partial-TDE disk}
Let us assume that the compact disk which forms after the TDE will have an outer radial scale at the circularisation radius of the stellar orbit\footnote{The value $2$ here is really $1+e$, where $e$ is the eccentricity of the incoming stellar orbit. This means that the disks formed in subsequent passages of the stellar remnant will have subtly different circularisation radii, something we neglect in this section as it has negligable impact on the disk accretion rate.} 
\begin{equation}
    r_c = 2 r_p = {2 \over \beta} r_T, 
\end{equation}
where the factor two here results from angular momentum conservation as the parabolic orbit is turned into a circular orbit. This results in an initial surface density scale of the accretion flow (likely reasonably accurate after $\sim 1$ viscous time) of  
\begin{multline}
    \Sigma_0 \simeq {M_{\rm disk} \over \pi R_{\rm out}^2} \simeq \left({\beta\over \beta_c}\right)^{2+q}  \left({f_d M_\star \beta_c^2 \over 8 \pi r_T^2}\right) \\ \simeq \left({\beta\over \beta_c}\right)^{2+q} \left({f_d M_\star^{5/3} \beta_c^2 \over 8\pi R_\star^2 M_\bullet^{2/3}}\right), 
\end{multline}
where in the final expression we have substituted for the canonical tidal radius $r_T = R_\star (M_\bullet/M_\star)^{1/3}$, and we define the parameter $0 \leq f_d \leq 1$ which represents what fraction of the stripped material which ultimately forms into a disk $M_{\rm disk} = f_d M_{\rm fb}$. 

The key result here is that we pick up {\it $\sim$ five powers} of impact parameter for the case of a partial TDE \citep[taking the][$q=3$ result]{Ryu+20a}, meaning if $\beta/\beta_c \simeq 1/2$ we immediately lose $\sim$ two orders of magnitude in surface density. After one viscous timescale the disk will continue to lose mass through its inner edge, while also expanding so as to conserve angular momentum. During this evolutionary phase the total angular momentum of the disk remains constant, meaning 
\begin{equation}
    J_{\rm disk}(t) \propto M_{\rm disk}(t) \sqrt{R_{\rm out}(t)} = {\rm constant}. 
\end{equation}
Meaning that at later times we have 
\begin{equation}
    \Sigma(t) \simeq {M_{\rm disk}(t) \over \pi R_{\rm out}^2(t)} \simeq \left({\beta\over \beta_c}\right)^{2+q} \left({f_d M_\star^{5/3} \beta_c^2 \over 8\pi R_\star^2 M_\bullet^{2/3}}\right) \left(1 - \Delta(t)\right)^5 , 
\end{equation}
where $\Delta (t)$ is the fraction of material which has been accreted up until time $t$.  This is given by 
\begin{equation}
    \Delta (t) = {1\over  M_{\rm disk}(0) } \int_{0}^t \dot M_{\rm H}(t') \, {\rm d}t' \simeq\left[1 - \left({t\over t_{\rm visc}}\right)^{1-n}\, \right], 
\end{equation}
where $\dot M_{\rm H}$ is the accretion rate across the horizon, and the final expression is approximately valid only for $t \gtrsim t_{\rm visc}$. The index $n$ (also the index at which the bolometric luminosity of the disk decays) depends on some precise assumptions about the turbulent transport in the disk, but is typically around $n \approx 1.2$ \citep[e.g.,][]{Cannizzo90}. This means for times large compared to the viscous timescale we have 
\begin{equation}
    \Sigma(t\gtrsim t_{\rm visc})  \simeq  \left({\beta\over \beta_c}\right)^{2+q} \left({f_d M_\star^{5/3} \over 8\pi R_\star^2 M_\bullet^{2/3}}\right) \left({t_{\rm visc} \over t}\right)^{5n-5} . 
\end{equation}
The viscous timescale in a TDE disk depends only on stellar properties 
\begin{multline}
    t_{\rm visc} = \alpha^{-1} \left({h\over r}\right)^{-2} \sqrt{r_c^3 \over GM_\bullet} = \alpha^{-1} \left({h\over r}\right)^{-2} \sqrt{8R_\star^3 \over \beta^3 GM_\star} \\ \equiv \beta^{-3/2} {\cal V} t_\star,
\end{multline}
where we have defined the dimensionless nuisance parameter ${\cal V} \equiv \alpha^{-1} \left({h / r}\right)^{-2} \gg 1$, and labeled $t_\star \equiv \sqrt{8R_\star^3 / GM_\star }$, which is the dynamical timescale of the star (up to order unity constants). We therefore have 
\begin{multline}
    \Sigma(t\gtrsim t_{\rm visc})  \simeq \left({\beta\over \beta_c}\right)^{(2q+19-15n)/2} \left({f_d M_\star^{5/3} \beta_c^2\over 8\pi R_\star^2 M_\bullet^{2/3}}\right)\\\left({ {\cal V}t_\star \over \beta_c^{3/2}t} \right)^{5n-5} .
\end{multline}
We see that the density suppression for a partial TDE disk goes like $\sim \beta^{7/2}$ at late times (for canonical choices of $n$ and $q$), which is slightly weaker than the early time suppression owing to the slightly longer evolutionary timescale expected for the larger initial disk. The suppression is still significant.

\subsection{Eddington ratios of partial TDEs}
The impact parameter dependence of the Eddington ratio of a partial TDE disk can be found from the simple constraints of total mass conservation. As we have a characteristic mass scale in the system ($M_{\rm disk}$) and characteristic timescale ($t_{\rm visc}$), we can state 
\begin{equation}
    \dot M_{\rm peak} \, t_{\rm visc} \sim M_{\rm disk},
\end{equation}
where we have neglected various order unity constants (the accretion disk of course does not accrete all of its matter in precisely one viscous timescale) which are all broadly irrelevant to the general point made here. The peak accretion rate of the disk then scales as 
\begin{equation}
    \dot M_{\rm peak} \sim M_{\rm disk}/t_{\rm visc} \sim \left({\beta\over \beta_c}\right)^{q+3/2} \, {f_d  \over \cal V} \sqrt{GM_\star^3 \beta_c^{3} \over 32 R_\star^3} , 
\end{equation}
where we again find an extremely strong dependence on the impact parameter. At late times ($t \gtrsim t_{\rm visc}$) we again have 
\begin{multline}
    \dot M(t\gtrsim t_{\rm visc}) \sim \dot M_{\rm peak} (t/t_{\rm visc})^{-n}  \\ \sim \left({\beta\over \beta_c}\right)^{(2q+3-3n)/2} \, \left[{f_d  \over \cal V} \sqrt{GM_\star^3 \beta_c^3\over 32 R_\star^3}\,\,\right]\left({ {\cal V } t_\star \over \beta_c^{3/2}t} \right)^{n} 
\end{multline}
which is a roughly $\sim \beta^{5/2}$ scaling (for our canonical model). Luminosity scalings are slightly less simple to determine, as the character of the accretion flow changes at the Eddington accretion rate ($\dot M_{\rm edd}\equiv L_{\rm edd}/\eta c^2$, with $\eta\sim 0.1$ the accretion efficiency) as energy advection thereafter plays an important role \citep{SS73}.  Above the Eddington limit the bolometric luminosity of the disk is effectively Eddington limited, scaling only logarithmically with an increase in mass accretion rate ($L_{\rm bol} \sim \dot M_{\rm edd}[1 + \ln(\dot M/\dot M_{\rm edd})]$, \citealt{SS73}), while below Eddington we expect $L_{\rm bol} \sim \dot M$ provided $t \gtrsim t_{\rm visc}$. 

One problem with interpreting the absolute values of the Eddington ratio at peak (in common with any time-dependent accretion model) is that one cannot determine the value of ${\cal V}$ from first principle arguments. In a real physical system it is effectively determined by the strength of magnetohydrodynamic turbulence in the disk, which is not something amenable to a simple analysis.  For the fiducial plots in this paper we take ${\cal V} = 1000$, a reasonable value (if one subscribes to the $\alpha$-viscosity phenomenology then this would be the value for $\alpha \sim h/r \sim 0.1$ which might be plausible for a TDE disk). We are therefore very fortunate that the accretion rate at late times shows such a weak dependence on this unknown $\dot M(t\gtrsim t_{\rm visc})\sim {\cal V}^{n-1} \sim {\cal V}^{1/5}$. Our most interesting results (those at late times) are therefore robust to this uncertainty\footnote{They are also completely insensitive to the eccentricity of the incoming stellar orbit, which causes only a $(1+e)^{3/10}$ scaling of the late time accretion rate, or a maximum change of $2^{3/10}\approx 1.23$. }.

{Another source of uncertainty is the precise value of the critical full-disruption impact parameter $\beta_c$, which is to leading order set by the density contrast of the incoming star}
\begin{equation}
    \beta_c \sim \left({\rho_{\star, \rm core}\over \rho_{\star, \rm avg}}\right)^{1/3} ,
\end{equation}
{where $\rho_{\star, \rm core}/ \rho_{\star, \rm avg}$ are the core/average density of the star \citep{Coughlin22b}. Fortunately however, once again this factor only enters as a weak power at late times $\beta_c^{3(1-n)/2} \sim (\rho_{\star, \rm core}/ \rho_{\star, \rm avg})^{-1/10}$ (assuming $n=1.2$), again leading to minimal uncertainty in our late-time results.   }

If the disks formed in tidal disruption events undergo the same state transitions as X-ray binary disks, then we would expect to see transitions in the disk when $L_{\rm bol}/L_{\rm edd} \sim f_{\rm tr}$, where $f_{\rm tr} \sim 0.01$. The time at which this transition occurs, which we call $t_{\rm tr}$ would scale with impact parameter as 
\begin{equation}
    t_{\rm tr} \sim \beta^{(2q + 3-3n)/2n} \sim \beta^{9/4}, \label{part}
\end{equation}
where the last scaling assumes our canonical model $n \simeq 1.2, q=3$.

\subsection{ The late-time UV plateau luminosity }
{In addition to thermal X-ray emission, t}he most clear signature of disk accretion in tidal disruption events is the late time flattening of the optical/UV light curves to a near-time independent state, typically occurring roughly $\sim 1$ year after discovery \citep{vanVelzen19, MumBalb20a, Mummery_et_al_2024, Guolo25c}. The near time-independence of this light curve phase can be understood as the cancellation of two competing effects, namely disk cooling (which would lower the optical/UV disk luminosity) and disk spreading (which would raise the optical/UV disk luminosity). A key property of this phase is that the amplitude is proportional to the black hole mass to the two-thirds power $L_{\rm plat} \propto M_\bullet^{2/3}$ \citep{Mummery_et_al_2024}, meaning that observations of this light curve phase can be used to constrain the black hole masses at the heart of these events. 

It is an important question to ask therefore how much these results are modified for partial TDEs, and whether or not the same scaling relationship can be applied. To answer this question we utilize the analytical results of \cite{Mummery_et_al_2024}, who show that the disk luminosity in the limit where the plateau luminosity originates from the outer radii of the disk scales as 
\begin{equation}
    L_{\rm plat} \sim r_c^{5/4} \, T_p \, t_{\rm visc}^{-(9n-10)/4} ,
\end{equation}
where $r_c \sim 1/\beta$ is the usual circularisation radius, the peak disk temperature scale reached during the evolution of the system is $T_p \sim \beta^{3/8} (M_{\rm disk}/M_\star)^{1/4}$, and $(9n - 10)/4 \simeq 1/5$ in the \cite{Cannizzo90} model. Note that we only show the relevant $\beta$-dependent scaling relationships here, the dependence on (e.g.,) stellar and black hole parameters are unchanged from that presented in \cite{Mummery_et_al_2024}. For partial TDEs we therefore have 
\begin{equation}
    L_{\rm plat} \sim \beta^{(27n+2q-37)/8} \sim \beta^{7/40},
\end{equation}
a weak dependence which lowers the amplitude from a full disruption, but not as strongly as might have been imagined. This is because although the reduced disk mass lowers the disk temperature (lowering the optical/UV flux), a partial TDE results in a larger accretion disk (raising the optical/UV flux). Therefore the weak dependence of the plateau luminosity on $\beta$ is another example of this fortuitous cancellation. This $\beta$ dependence is typically weaker than the dependence on all other parameters (stellar mass, inclination, etc.) and therefore the scaling relationship derived for full disruptions still broadly applies for partial disruptions. This only holds, however, if the time to transition to a hard state (which may well modify the disk structure substantially) is longer than the time for the plateau to be revealed, typically $\Delta t_{\rm obs} \sim 1-2$ years. As we demonstrate in the next section this may well not be the case, at least for some partial TDE systems, meaning this scaling relationship should be used carefully on a case-by-case basis. 

\subsection{The transition to quiescence }
{In addition to undergoing a transition to a harder (coronal dominated) X-ray state at $\sim 1\%$ of the Eddington luminosity, X-ray binaries are further known to transition to quiescent states at even lower Eddington ratios. The same final transition has been observed in TDEs, including AT2018fyk in \cite{Wevers21}, where the second transition occurred at an Eddington ratio of $f_{\rm edd} \sim 10^{-3}$, i.e., roughly one order of magnitude lower than for the original soft-to-hard state transition.   }

{We note an interesting result here, that if one constrains the time to fall to the hard state transitional scale observationally ($t_{\rm hard}$), then as none of the other parameters in the problem can vary, the time to fall to quiescence ($t_{\rm Q}$) is just }
\begin{equation}
    t_Q \approx t_{\rm hard}\left({ f_{\rm hard} \over f_Q }\right)^{1/n}\approx 7 t_{\rm hard},
\end{equation}
{where $f_{\rm hard}/f_Q$ are the hard state/quiescence transitional Eddington ratios respectively, and the final expression is an estimate for the TDE case (admittedly built on a sample of 1) where the required fall in Eddington ratio is an order of magnitude between the two transitional scales. This highlights that when the hard state transitional time is short (as in for partial TDEs), one has a very real possibility of observationally probing the transition to a quiescent state (as was seen in AT2018fyk \citealt{Wevers21}, on two occasions \citealt{Pasham24fyk}). For full disruptions the original hard state transitional time can be $\sim 1000$'s of days (as we shall now discuss), meaning the disk may well be in a bright state (above the quiescent transition) for decades.  }

\subsection{Results for full disruptions $\beta \geq \beta_c$}
For full disruptions, with $\beta\geq\beta_c$, the results of this analysis are modified. We shall assume that one half of the incoming star remains bound for all $\beta\geq\beta_c$, meaning the only $\beta$ dependence of the emission enters through the circularisation radius $r_c \sim 1/\beta$, and associated evolutionary timescale of the disk $t_{\rm visc} \sim r_c^{3/2} \sim \beta^{-3/2}$. The scaling arguments carried out above can be simply repeated, leading to (where the final scalings assume $n=6/5$)
\begin{align}
    \Sigma_0 &\sim \beta^2, \\
    \Sigma(t \gtrsim t_{\rm visc}) &\sim \beta^{(19-15n)/2} \sim \beta^{1/2}, \\
    \dot M_{\rm peak} &\sim M_{\rm disk}/t_{\rm visc} \sim \beta^{3/2},  \\
    \dot M(t \gtrsim t_{\rm visc}) &\sim \beta^{3(1-n)/2} \sim \beta^{-3/10},  \\
    t_{\rm tr} &\sim \beta^{(3-3n)/2n} \sim \beta^{-1/4}, \\
    L_{\rm plat} &\sim \beta^{-3/5} . \label{full}
\end{align}
We note that the transitional timescale shows a weak, {\it negative} dependence on $\beta$ in this limit, implying that disks formed in $\beta=\beta_c$ tidal disruption events stay above a certain Eddington ratio scale for the longest. The plateau luminosity also decreases with increasing $\beta$, owing to the more compact disks formed in deep disruptions. 

\section{Numerical confirmation}\label{num}
To test the scaling arguments put forward in this paper we use the {\tt FitTeD} code \citep{mummery2024fitted}, which solves the time-dependent relativistic accretion disk equations for the disk temperature, and then computes the integrated X-ray  (defined in this work as the integrated 0.3–10 keV disk luminosity), and bolometric luminosity of the disk systems. For canonical system parameters we take $M_\star = 0.8 M_\odot$, ${\cal V} = 1000$, $M_\bullet = 10^6 M_\odot$, $a_\bullet = 0$, $i = 45^\circ$. {All numerical solutions in this section take $q=3$ and $\beta_c = 2$ (a plausible value for a low mass zero age main sequence star \citealt{Coughlin22b}).} We use a stellar mass-radius relationship of the form  $R_\star = 0.8 R_\odot (M_\star/M_\odot)^{0.93}$, which is taken from \citep{Ryu+20a}, and take $f_d = 0.1$ as suggested by simulations \citep{Bonnerot21}. The disk is assumed to form as a ring at the $\beta$-dependent circularisation radius discussed above, and then evolves following the standard time-dependent general relativistic thin-disk equations \citep{Balbus17}. The solutions computed in {\tt FitTeD} have a bolometric decay index $n = 4/3$. 

\begin{figure}
    \centering
    \includegraphics[width=0.99\linewidth]{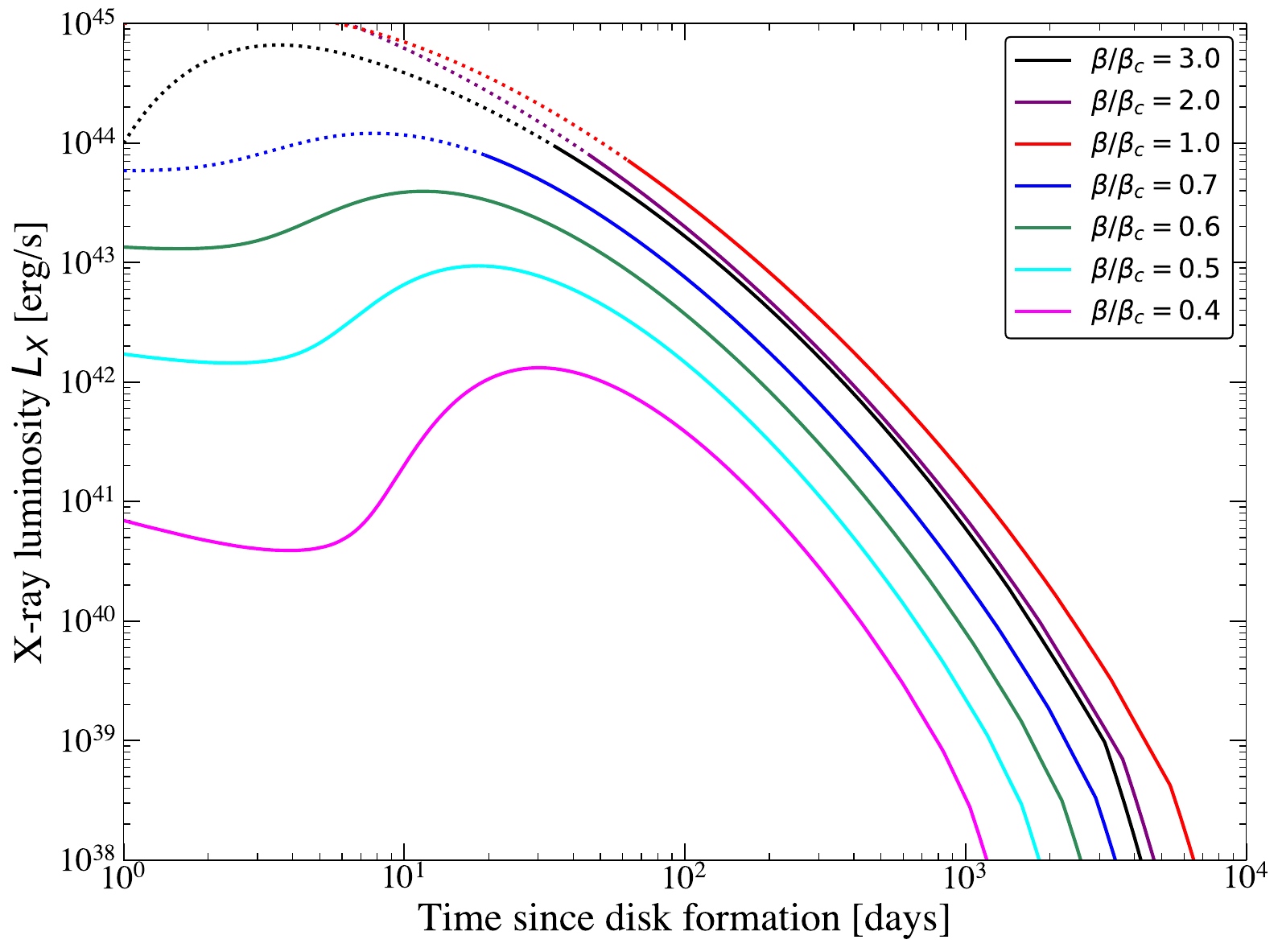}
    \includegraphics[width=0.99\linewidth]{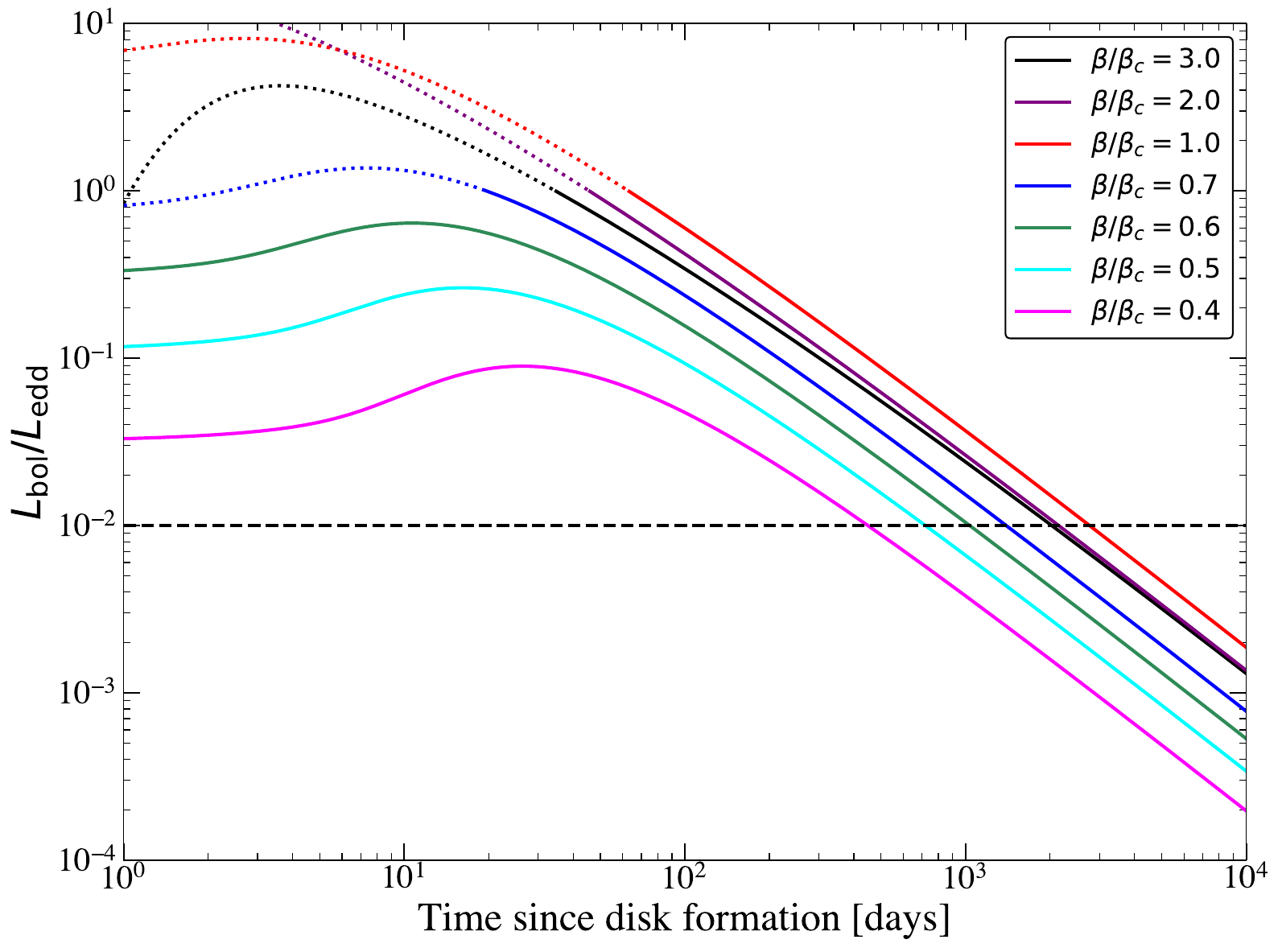}
    \caption{{\bf Upper:} the X-ray light curves (defined here as the integrated luminosity between photon energies of 0.3 and 10 keV) for fiducial stellar and black hole parameters (see text), and different values of the impact parameter $\beta$ (displayed on plot). Even weak disruptions $\beta/\beta_c \sim 1/2$, where only $\sim 10\%$ of the mass is stripped off the star can produce observable X-ray flares. {\bf Lower:} the Eddington luminosity ratio of the disk systems, as a function of impact parameter $\beta/\beta_c$. By a horizontal dashed line we display an Eddington ratio of $1\%$, highlighting the different timescales at which the partial and full disruptions cross this transitional scale.  }
    \label{fig:partial+full}
\end{figure}

\begin{figure}
    \centering
    \includegraphics[width=0.99\linewidth]{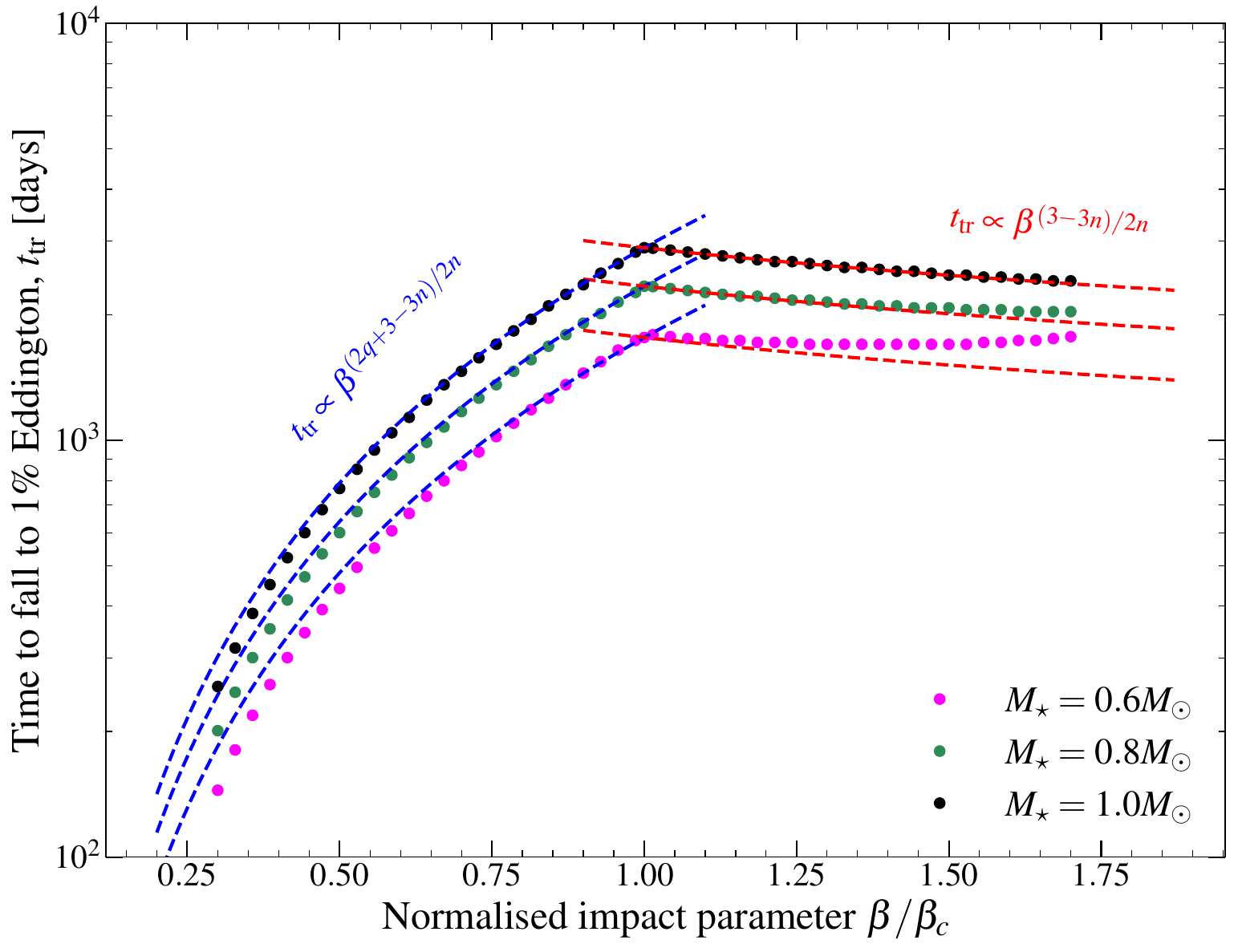}
    \caption{ The time taken (in days) for the accretion flow to reach an Eddington ratio of $1\%$, from the time of first disk formation, as a function of {normalised} impact parameter $\beta/\beta_c$. The dashed lines show the analytical scaling argument put forward in this work. For sufficiently small impact parameters $\beta/\beta_c \sim 1/2$ state transitions could occur on timescales of hundreds of days. Deviations from the predicted scaling relationship at very low $\beta/\beta_c$ occur as the disk has not reached the $t \gg t_{\rm visc}$ asymptotic regime when $1\%$ Eddington is reached. The deviation at high $\beta/\beta_c$ results from the relativistic formation radius of the disk, which leads to modifications from the strictly Newtonian analytical scaling arguments used in the text. }
    \label{fig:transition}
\end{figure}

\begin{figure}
    \centering
    \includegraphics[width=0.99\linewidth]{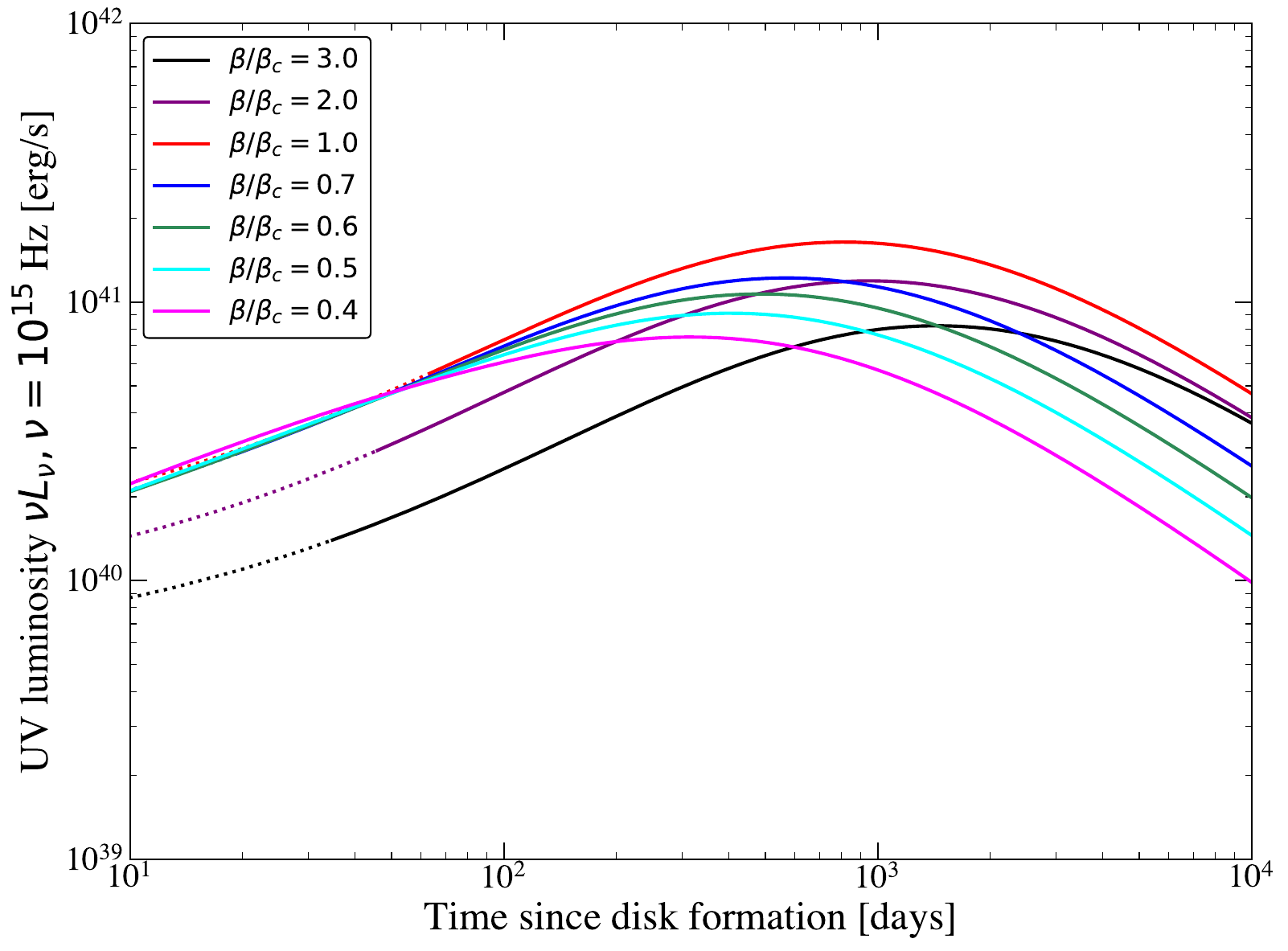}
    \caption{The light curves of the accretion flows shown in Figure \ref{fig:partial+full}, as observed at UV frequencies ($\nu = 10^{15}$ Hz). We see that the UV plateau luminosity roughly $\sim 1000$ days post disruption shows only a weak dependence on $\beta/\beta_c$, with $\beta/\beta_c = 1$ disruptions brightest (as can be understood analytically). At significantly later times ($\sim$ decades) partial disruptions show a more pronounced $\beta$ dependence, although the disks are likely to have undergone a state transitions by this point, which may well modify their behavior.  }
    \label{fig:uv}
\end{figure}

In Figure \ref{fig:partial+full} we display the X-ray luminosity (upper panel) and Eddington luminosity ratio (lower panel) of these canonical disk systems, for different values of the impact parameter $\beta$. The key points to highlight here are (i) even weak disruptions $\beta/\beta_c \sim 1/2$, where only $\sim 10\%$ of the mass is stripped off the star, produce observable X-ray flares $L_{X, {\rm peak}} \gtrsim 10^{42}$ erg/s; (ii) increasing $\beta/\beta_c > 1$ leads to disks which are more super-Eddington at early times, but which transition through the relevant sub-Eddington transitional scales earlier than $\beta/\beta_c = 1$ systems owing to their quicker evolution; (iii) partial disruptions $\beta/\beta_c < 1$ transition through lower Eddington ratios much more rapidly than full disruptions. All of these results are precisely in keeping with the analytical scaling arguments presented above. We note that for some early times the disk solutions found by the {\tt FitTeD} code are formally super-Eddington $L_{\rm bol} > L_{\rm edd}$, where the governing assumptions of the model begin to break down. Periods of time where this is true are displayed by dotted curves, and should not be interpreted literally.  

If one takes the state-transition Eddington ratio to be $f_{\rm edd} = 0.01$, then it is possible to compute from the {\tt FitTeD} disk solutions the time taken for a given solution to pass through the point at which  $L_{\rm bol} = f_{\rm edd} L_{\rm edd}$. As in earlier sections we call the time to reach this point $t_{\rm tr}$, which is plotted as a function of impact parameter (for three different stellar masses denoted on the plot) in Figure \ref{fig:transition}. Except for stellar mass all other parameters are the same as in Figure \ref{fig:partial+full}. By solid points we display the numerical results, and with dashed curves we display the scaling relationships derived above, for partial disruptions in blue (equation \ref{part}), and for full disruptions in red (equations \ref{full}). Deviations from the predicted scaling relationship (the transitions happen even quicker than predicted) occur at very low $\beta/\beta_c$  as the disk has not reached the $t \gg t_{\rm visc}$ asymptotic regime when $1\%$ Eddington is reached, but the general agreement is excellent. We see that a typical full disruption may remain in a softer state for $t \gtrsim 1000$'s of days \citep[in good agreement with the observed evolution of ASASSN-14li, for example][]{Bright18}, while partial TDEs can transition in only a few hundred days.

These numerical results allow us to calibrate the various order unity constants left out of the analytical scaling arguments, and we find that 
\begin{multline}\label{transition_time}
    {t_{\rm tr} \over 4000\, {\rm days}} \approx f(\beta/\beta_c) \left[{{\cal V} \over 1000} \beta_c^{-3/2} \sqrt{R_\star^3 M_\odot \over R_\odot^3 M_\star} \,\,\right]^{(n-1)/n}\\\left[\left({\eta(a)\over \eta(0)}\right) \left({f_d\over 0.1}\right) \left({M_\star \over M_\odot}\right)  \left({0.01 \over f_{\rm tr}}\right) \left({10^6 M_\odot \over M_\bullet}\right)\right]^{1/n} ,
\end{multline}
where 
\begin{equation}
    f(\beta/\beta_c) = \begin{cases}
        (\beta/\beta_c)^{(2q+3-3n)/2n} , \quad \beta/\beta_c \leq 1,  \\
        \\
        (\beta/\beta_c)^{(3-3n)/2n} , \quad\quad\,\, \beta/\beta_c > 1. 
    \end{cases}
\end{equation}
Here $\eta(a)$ is the usual spin-dependent accretion efficiency, equal to $\eta = 1 - (1-{2GM_\bullet/ 3r_Ic^2})^{1/2}$, where $r_I$ is the spin-dependent ISCO radius \citep[e.g.,][]{Bardeen72}. We note that the precise numerical values of (e.g.,) the X-ray luminosity amplitudes shown in Figure \ref{fig:partial+full} should be taken as only illustrative, as the X-ray luminosity is a sensitive (typically exponential) function of various parameters, including the disk-observer inclination angle and black hole spin. Further, once the disk has transitioned through to the hard state, the subsequent luminosity evolution will be strongly modified from what is shown here, as the non-thermal X-ray emission begins to dominate. Attempts have been made to model the disk evolution beyond the hard-state transition \citep{MumBalb21b}, but the final transition to the quiescent state (which can occur rapidly in X-ray binaries), is poorly understood. 

We note that the $t_{\rm tr} \sim 1/M_\bullet^{1/n}$ scaling derived here can be thought of as an additional test of this general framework, with TDEs around more massive black holes expected to undergo state transitions more rapidly \citep[an earlier, similar,  argument along these lines was put forward by][]{MumBalb21b}. 

Finally, we examine the properties of the late-time UV luminosity in these systems. In Figure \ref{fig:uv} we display the UV luminosity which would be observed at a (rest frame) frequency of $\nu = 10^{15}$ Hz, for each of the disk systems shown in Figure \ref{fig:partial+full}. We see, as predicted analytically, that the UV plateau luminosity roughly $\sim 1000$ days post disruption shows only a weak dependence on $\beta$, with $\beta/\beta_c = 1$ disruptions brightest. At significantly later times ($\sim$ decades) partial disruptions show a more pronounced $\beta$ dependence, although the disks are likely to have undergone a state transitions by this point, which may well modify their behavior.  

We know of only two repeating TDEs with observational baselines long enough to detect a UV plateau, namely AT2018fyk \citep{Wevers21, Mummery_et_al_2024} and AT2022dbl \citep{Arcavi22TNS, MummeryVV25}. Both of these systems had black hole masses inferred from their late time UV observations in keeping with other estimates using galactic scaling relationships, in support of the analysis put forward here.  

\section{A case study of AT2018fyk}
\begin{figure}
    \centering
    \includegraphics[width=0.99\linewidth]{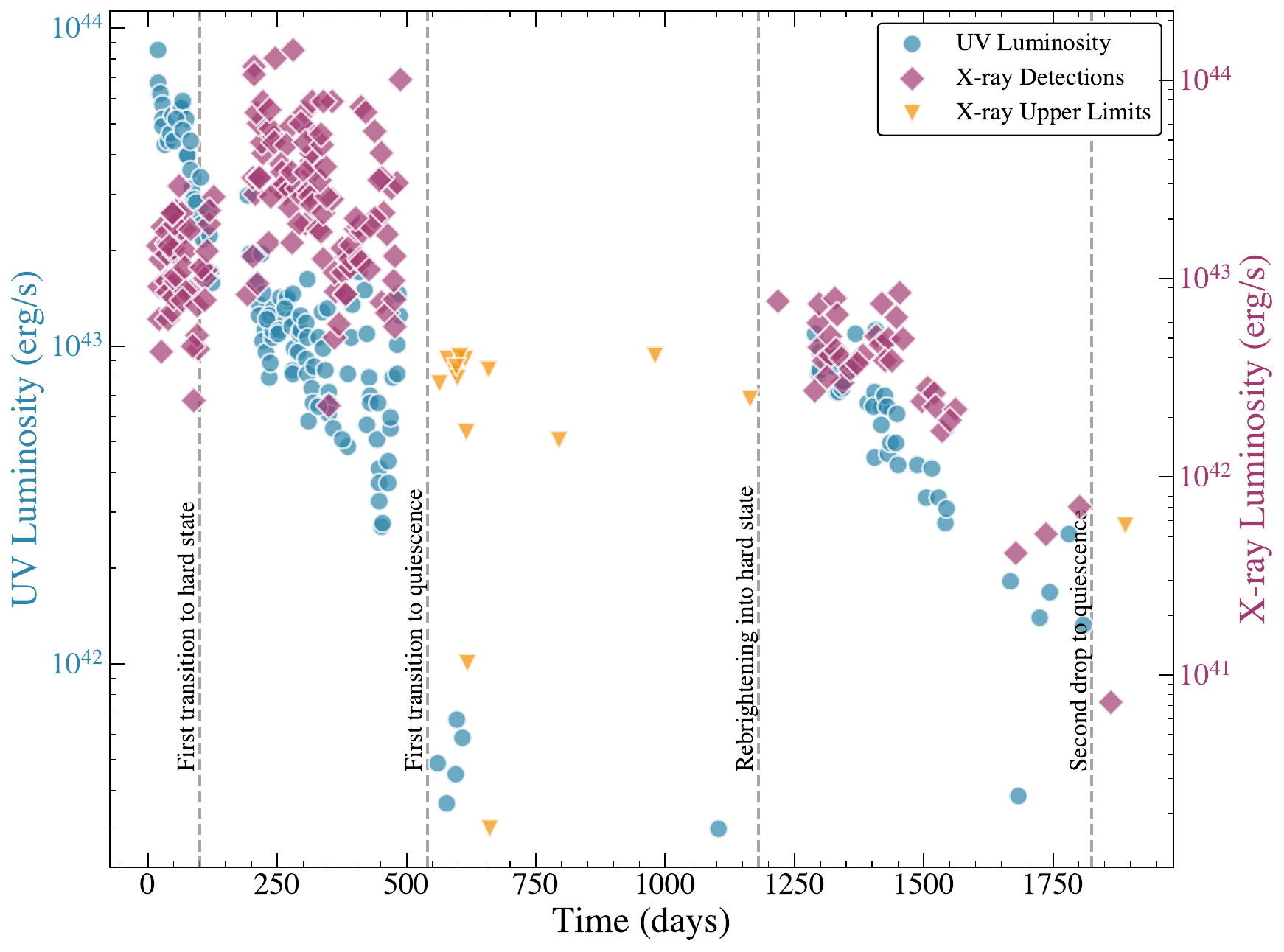}
    \caption{The multi wavelength light curves of AT2018fyk, data from \citealt{Pasham24fyk}. The evolving UV luminosity is shown by blue points (left vertical axis), the evolving X-ray luminosity by purple diamonds (right vertical axis), while X-ray upper limits are shown by orange inverted triangles. The rough timescales at which different state transitions occur are shown by vertical dashed lines.  }
    \label{fig:18fyk}
\end{figure}

{AT2018fyk is perhaps the best observationally constrained repeating partial TDE \citep{Wevers21, Wevers23fyk, Pasham24fyk}, with a bright UV plateau \citep{Wevers21}, an observed transition from soft to hard X-ray state at $t \sim 100$ days post disruption \citep{Wevers21}, a further transition to the quiescent state at $t \sim 580$ days post disruption \citep{Wevers21}, a re-brightening into a hard accretion state at $t \sim 1180$ days post {\it initial}  disruption \citep{Wevers23fyk} and a final transition to a quiescent state at $t \sim 1825$ days post {initial} disruption \citep{Pasham24fyk}. This source therefore represents a good test of the framework put forward in this paper. Ideally, one would fit the multi wavelength observations of AT2018fyk with a full time-dependent disk model which includes transitions between different states. Such an analysis, while of great interest, lies beyond the scope of this current work, and we consider instead characteristic luminosity and time scales of the system. 

The velocity dispersion of AT2018fyk is $\sigma \approx 158$ km/s, implying a black hole mass of $\log_{10} M_\bullet/M_\odot = 7.7 \pm 0.4$ \citep{Wevers21}. We begin with the plateau luminosity, which for the \cite{Mummery_et_al_2024} scaling relationship and this mass would be }
\begin{equation}
    \log_{10} \left({L_{\rm UV, plat} \over {\rm erg/s}} \left({\beta \over \beta_c}\right)^{-7/40}  \right) = \left[42.7 \pm 0.2_{\rm mass} \pm 0.5_{\rm sys}\right] .
\end{equation}
{In this expression there is an uncertainty in the plateau luminosity from the uncertainty in the mass ($\pm 0.2$ dex) and from the intrinsic scatter in the relationship ($\pm 0.5$ dex). The impact parameter factor is only $\sim 0.9$ for $\beta = \beta_c/2$. This luminosity scale is entirely consistent with observed UV luminosity of AT2018fyk after $\sim 1$ year (Figure \ref{fig:18fyk}), and when the disk re-brightens into the hard state at $t \sim 1300$ days. 

We next ask if the initial hard state transition timescale makes sense for this source. Assuming otherwise standard parameters, namely a one solar mass star which forms into a disk (a high mass star makes sense for a high mass black hole such as AT2018fyk), one finds a transitional time to reach $1\%$ Eddington  (following equation \ref{transition_time}, which was calibrated with $q=3$ and $n=4/3$) of} \begin{equation}
    t_{\rm tr} \approx 90\, {\rm days} \, \left({2\beta \over \beta_c }\right)^{15/8} \eta_{-1}^{3/4} f_{d, -1}^{3/4} \alpha_{-1}^{1/4}\theta_{-1}^{1/2} m_\star^{5/8} r_\star^{3/8}M_{\bullet, \sigma}^{-3/4} 
     ,
\end{equation}
Here, $m_\star \equiv M_\star/M_\odot$, $r_\star \equiv R_\star/R_\odot$, and $\theta \equiv (h/r)$. A subscript $-1$ indicates that a value has been normalized by 0.1.
{We have taken $\log_{10} M_\bullet/M_\odot = 7.7 \equiv M_{\sigma}$ from the velocity dispersion, $M_{\bullet, \sigma} \equiv M_\bullet/M_\sigma$, and normalized by an initial impact  parameter $\beta = 0.5 \beta_c$. Of course, this time would systematically shift for different assumed transitional Eddington ratio scales, different assumptions about the turbulent stress, a different black hole mass consistent with the $M_\bullet-\sigma$ value, a different fraction of the stellar debris which entered the disk, etc., but it is clearly  consistent with the observed behavior (Figure \ref{fig:18fyk}). }

{We note that the peak X-ray luminosity scale of AT2018fyk in the hard state is $L_X \sim 10^{44}$ erg/s, precisely at the maximum scale derived in \cite{Mum21} for a large Comptonisation fraction and a high black hole mass (their Figure 8). }

{The time taken for the first drop into quiescence then follows from the discussion above, namely we use the fact that the observed Eddington ratio as the quiescent state was entered was roughly a factor $\sim 10$ lower than when the hard state was entered \citep{Wevers21}. The time at which this quiescent transition would be expected is then  }
\begin{multline}
    t_Q \approx t_{\rm hard} \left({ f_{\rm hard} \over f_Q }\right)^{1/n} \\ \approx 600\, {\rm days} \, \left({2\beta \over \beta_c }\right)^{15/8} \, \left({ f_{\rm hard} \over 10 f_Q }\right)^{3/4} \\ \times \eta_{-1}^{3/4} f_{d, -1}^{3/4} \alpha_{-1}^{1/4}\theta_{-1}^{1/2} m_\star^{5/8} r_\star^{3/8}M_{\bullet, \sigma}^{-3/4}  ,
\end{multline}
{where, once again, the exact value here is systematically uncertain for all the same reasons as the initial hard state transitional time was, but is certainly broadly consistent with what was observed (Fig. \ref{fig:18fyk}).   }

{The system then re-brightens, presumably as a result of more material being added to the disk from the returning stellar remnant. Once again, the UV luminosity reaches a scale of $L_{\rm UV} \sim 10^{42.5}$ erg/s, entirely as expected for direct accretion disk emission from a black hole of this mass scale (the plateau luminosity is weakly dependent on disk mass, which may well be different for this second flare).   }

{The X-ray spectrum of AT2018fyk on this second flare is immediately detected in the hard state \citep{Wevers23fyk}. This suggests that, indeed, less mass was stripped on this second passage meaning the peak Eddington ratio of this replenished disk was only just above/at the scale of $\sim 1\%$ Eddington. Accretion disk evolution is however self-similar (a result used throughout this paper), and so assuming that the saturated state of the turbulent disk stress was similar in both flares, and that the Eddington ratio drop required for the hard-to-quiescent transition was the same in both flares, we have  }
\begin{equation}
    t_Q^{(2)} - t_{\rm hard}^{(2)} \approx t_Q^{(1)} - t_{\rm hard}^{(1)} \approx 510\,{\rm days}.
\end{equation}
{Again, this is entirely consistent with what is observed, given the order one uncertainties and systematics. }

{Naturally, the simple time-scale and luminosity-scale analysis performed in this section should be extended with full disk modeling in future works, but AT2018fyk is certainly consistent with the framework put forward in this paper. }

\section{The next time round: flares from disk-remnant collisions?}\label{theory2}
A partially disrupted star will continue on its orbit, and thereafter return to roughly the same pericentre. On its second lap of the black hole it shall encounter an accretion flow of material, and it is natural to ask whether an interaction between the returning star and the pre-existing disk could produce any observable signatures. Recently \cite{Linial2023} have developed a star-disk collision model aiming to describe so-called quasi-periodic eruptions \citep[e.g.,][]{Miniutti2019, Giustini2020, Arcodia2021, Arcodia2024a, Guolo2024, Nicholl24, Chakraborty25}, and we utilize their framework here\footnote{We note that the original Linial \& Metzger framework is no longer the preferred model for collision-induced QPEs, with various groups \citep{Yao24, Mummery25QPE, Linial25qpe2} arguing that an extended debris stream (stripped from the star over repeated disk-crossings) is required to reproduce QPE observables within the collisional framework, particularly for a group of long period QPEs in TDEs with high radiated energy budgets \citep{Mummery25QPE}}.  We begin by estimating the kinetic energy imparted into the disk material by a collision between the disk and the returning remnant. The velocity of the remnant when it returns to pericentre will be that of an elliptical orbit with pericentre $r_T/\beta$ and eccentricity $e_{\rm orb}$ (the remnant must be on an eccentric orbit for it to return)
\begin{equation}
    v \approx \sqrt{GM_\bullet \beta(1+e_{\rm orb}) \over r_T } \approx \sqrt{GM_\bullet^{2/3} M_\star^{1/3} \beta (1+e_{\rm orb})  \over R_\star }, 
\end{equation}
and will collide with a disk surface density $\Sigma$, given above. The amount of mass shock heated and ejected in this collision is 
\begin{multline}
    M_{\rm ej} \approx 2 \pi \left(R_\star^{\rm ret}\right)^2 \, \Sigma_{\rm disk}(t_{\rm ret}) \\
    \approx \left({R^{\rm ret}_\star \over R_\star}\right)^2  \left({\beta\over\beta_c}\right)^{2+q} \left({f_d M_\star^{5/3} \beta_c^2 \over  4  M_\bullet^{2/3}}\right) \left(1 - \Delta(t_{\rm ret})\right)^5 ,
\end{multline}
where $R_\star^{\rm ret}$ is the radius of the star {\it once it has returned after its first disruption}, and $t_{\rm ret}$ is the time taken for the stellar remnant to return to pericentre. For notational ease let us define two unknowns, the first being \begin{equation}
    {\cal A} \equiv {M_{\rm disk}(t_{\rm ret}) \over M_{\rm disk}(0)} \approx \left({ {\cal V}t_\star \over t_{\rm ret}\beta^{3/2} } \right)^{5n-5} ,
\end{equation} 
the fraction of the initial disk material which remains upon the remnant's return (this expression is only valid for ${\cal A}\leq 1$, and ${\cal A}$ should be taken equal to one otherwise). {Note that this factor depends both on the time taken for the remnant to return to pericentre, and also the viscous timescale of the initial partial-TDE disk. Through this second parameter it does implicitly depend on the original impact factor ${\cal A}={\cal A}(\beta)$.} {Let us also denote the radius ratio }
\begin{equation}
    {\cal R} \equiv \left({R^{\rm ret}_\star \over R_\star}\right) = {\cal R}(\beta/\beta_c, q), 
\end{equation}
{which will presumably depend both on the amount of mass lost from the star on the previous passage and the depth of the first tidal interaction. An important limit is that ${\cal R}= 0$ for $\beta/\beta_c \geq 1$ (full disruptions).} In this case the kinetic energy in the shocked gas is  
\begin{align}
    E_{\rm kin} &= {1\over 2} M_{\rm ej} v^2, \\
    &\approx  \left({\beta\over \beta_c}\right)^{3+q} {\cal R}^2 (1+e_{\rm orb})\, f_d \, {\cal A}^5\,\beta_c^3\, {G  M_\star^2 \over 4 R_\star},  \\ &\approx 10^{48}\, {\rm erg} \, \, \left({\beta\over \beta_c}\right)^{3+q} {\cal R}^2 (1+e_{\rm orb})\, f_d \, {\cal A}^5\,\beta_c^3\, \nonumber\\ &\quad\quad\quad \left({M_\star \over M_\odot}\right)^2 \left({R_\star \over R_\odot}\right)^{-1}  .
\end{align}
The central expression here highlights the interesting result that the kinetic energy in the collision between the returning stellar remnant and the disk formed from its previous passage is set in scale by its original gravitational binding energy $E_{
\rm kin} \propto GM_\star^2/R_\star$ multiplied by a $\beta-$dependent prefactor, and is independent of the black hole mass at the center of the event. The ejected mass in this limit is given by 
\begin{multline}
    M_{\rm ej} \approx 3 \times 10^{-5}\,  M_\odot \, f_d {\cal A}^5 \,  \left({\beta\over \beta_c}\right)^{2+q}{\cal R}^2 \beta_c^2 \\ \left({M_\star \over M_\odot}\right)^{5/3} \left({M_\bullet \over 10^6 M_\odot}\right)^{-2/3} ,
\end{multline}
which will result in an evolving optical depth of the cloud of 
\begin{equation}
    \tau(t) \approx {M_{\rm ej}\kappa_{\rm es} \over 4 \pi R_{\rm ej}^2} ,
\end{equation}
where we shall assume that the radius of the ejecta will evolve ballistically (as $R_{\rm ej} \sim v t$). Treating this system like the shock-cooling emission following a supernovae, photons will escape when the optical depth drops to $\tau(t_{\rm flare}) \sim c/v$, or after a time 
\begin{align}
    t_{\rm flare} &\approx \sqrt{M_{\rm ej}\kappa_{\rm es} \over 4 \pi v c}, \\ 
    &\approx 0.8 \, {\rm hr} \, f_d^{1/2} {\cal A}^{5/2} (1 + e_{\rm orb})^{-1/2} \, \left({\beta\over\beta_c}\right)^{(3+2q)/4} \beta_c^{1/4} {\cal R} \nonumber  \\ &\quad\quad\quad\left({M_\star \over M_\odot}\right)^{3/4} \left({R_\star \over R_\odot}\right)^{1/4} \left({M_\bullet \over 10^6 M_\odot}\right)^{-1/2} .
\end{align}
Up to order unity constants this will set the time-scale for the rise and decay of any emission produced by this collision. This timescale also sets the volume expansion factor of the gas before any radiation is emitted, and therefore the scale of the adiabatic losses which reduce the available energy from the initial kinetic energy of the gas cloud. The initial density of the post-shocked cloud is given by $\rho_{\rm sh} \approx \rho_{\rm disk} (\gamma + 1)/(\gamma - 1)  = 7 \rho_{\rm disk}$, where $\rho_{\rm disk}$ is the density of the gas in the disk (this is just a statement of the strong shock conditions which are valid here as the remnant is moving highly supersonically with respect to the disk fluid $v^2/c_s^2 \sim 2(h/r)^{-2} \gg 1$), and we assume a radiation pressure dominated equation of state ($\gamma = 4/3$).  This means that the width of the ejecta is initially $w \approx h/7$ where $h$ is the scale height of the disk.  The size of the ejecta once it begins to emit photons is $R_{\rm ej} \approx v t_{\rm flare}$, and so the energy available to be radiated from the cloud is $E_{\rm rad} \approx E_{\rm kin} (3V_0/4\pi R_{\rm ej}^3)^{\gamma - 1} $, where $V_0 \approx \pi (R_\star^{\rm ret})^2 w$, or explicitly 
\begin{equation}
E_{\rm rad} \approx E_{\rm kin}  \left({3 (R_\star^{\rm ret})^2 h \over 28  v^3 t_{\rm flare}^3 }\right)^{1/3} , 
\end{equation}
which once simplified becomes 
\begin{multline}
E_{\rm rad} \approx 9\times10^{45} \, {\rm erg} \, f_d^{1/2} {\cal A}^{5/2} \,\left({h\over r}\right)^{1/3} \left({\beta\over \beta_c}\right)^{(17+6q)/12} \beta_c^{17/12}  \\ {\cal R}^{5/3} (1+e_{\rm orb})^{3/4}  \left({M_\star \over M_\odot}\right)^{35/36} \left({R_\star \over R_\odot}\right)^{1/4} \left({M_\bullet \over 10^6 M_\odot}\right)^{5/18} .
\end{multline}
We see that at this point we have our first ``unknown'' from the disk, namely the disk scale height at the collision radius $h/r$. Fortunately this enters only as a weak power. 

This energy budget remaining in the radiation field, combined with the flare timescale, implies a peak luminosity of 
\begin{align}
    L_{\rm flare} &\approx {E_{\rm rad}\over t_{\rm flare}} , \\
    &\approx 3\times10^{42} \, {\rm erg/s}\,\left({h\over r}\right)^{1/3} \left({\beta\over\beta_c}\right)^{2/3} \beta_c^{7/6} {\cal R}^{2/3} \nonumber \\ &\quad\quad  (1+e_{\rm orb})^{5/4}\left({M_\star \over M_\odot}\right)^{2/9}  \left({M_\bullet \over 10^6 M_\odot}\right)^{7/9} ,
\end{align}
{which is independent of how the partial disruption mass stripping depends on $\beta$ (i.e., the index $q$).} We conclude that, unfortunately, it is unlikely that a $\sim 0.1-1$ hour flare of relatively weak luminosity $L_{\rm flare} \lesssim 10^{42}$ erg/s will be detectable, especially given the uncertainty in the return time of the stellar remnant. The two ways around this appear to be for larger black hole masses ($L_{\rm flare} \propto M_\bullet^{7/9}$), although this will further reduce the duration of the flare ($t_{\rm flare} \propto M_\bullet^{-1/2}$), or for very large mass stars ($L_{\rm flare} \propto M_\star^{2/9}, t_{\rm flare}\propto M_\star^{3/4}$). For completeness, it is worth determining the potential observational signatures of such a flare. The effective blackbody temperature of this radiation is  
\begin{multline}
k T_{\rm BB} =  k \left({c u_\gamma \over 4\sigma}\right)^{1/4} \approx k \left({3 c E_{\rm rad} \over 16 \pi \sigma R_{\rm ej}^3}\right)^{1/4} \\ = k \left({3 c E_{\rm rad} \over 16 \pi \sigma v^3 t_{\rm flare}^3}\right)^{1/4} , 
\end{multline}
where $u_\gamma$ is the radiation energy density, and $\sigma$ is the Stefan-Boltzmann constant. This evaluates to 
\begin{multline}
k T_{\rm BB} \approx 7 \, {\rm eV} \, \, f_d^{-1/4} {\cal A}^{-5/4} \left({h\over r}\right)^{1/12} \, \left({\beta\over \beta_c}\right)^{-(7+3q)/12} \\ \beta_c^{-1/2}{\cal R}^{-1/3} \,  \left({M_\star \over M_\odot}\right)^{-4/9} \left({R_\star \over R_{\odot}}\right)^{1/4}  \left({ M_\bullet \over 10^6 M_\odot }\right)^{7/36} . 
\end{multline}
Note that this temperature formally diverges in the limit $\beta \to 0$, which is (obviously) unphysical. This is irrelevant however (this is merely a formal limit), as the flare duration in this limits vanishes $t_{\rm flare} \to 0$. 

This temperature is, however, only the temperature which would be observed in the case that the photon field is able to reach thermodynamic equilibrium with the expanding gas. To test whether this is the case we compute \citep[following][]{Linial2023}
\begin{equation}
\eta \equiv {n_{\rm BB} (T_{\rm sh}) \over t_{\rm cross} \, \dot n_{{\rm ff}, \gamma} (\rho_{\rm sh}, T_{\rm sh})} .
\end{equation}
Here $\dot n_{{\rm ff}, \gamma}\propto \rho^{2}T^{-1/2}$ is the photon production rate of free-free Bremsstrahlung,  $\rho_{\rm sh} = 7\Sigma_{\rm disk}/(2h)$ is the density of the gas post-shock,  $T_{\rm sh} = (3c \rho_{\rm sh}v^2 / 4\sigma)^{1/4}$ is the effective blackbody temperature of this shocked gas, and $t_{\rm cross} = h/(7v)$ is the time it takes for the remnant to cross the shocked gas. The denominator therefore represents the number of photons which can be produced during the period of high-density post shock. The numerator $n_{\rm BB}(T)\approx 4\sigma T^3 / 3kc$ represents the number of photons which would have to be produced by Bremsstrahlung for the gas to reach thermodynamic equilibrium. Therefore if $\eta > 1$ thermodynamic equilibrium cannot be established, and photons will generally escape with harder energies than $kT_{\rm BB}$. Evaluating this criteria we find 
\begin{multline}
\eta \approx 0.2 \, {f_d^{-9/8}}{\cal A}^{-45/8} \left({\beta\over\beta_c}\right)^{-(9q+8)/8} \left({h\over r}\right)^{1/8} \left({M_\bullet \over 10^6 M_\odot}\right)^{41/24} \\ (1+e_{\rm orb})^{11/8} \beta_c^{13/4} \left({M_\star \over M_\odot }\right)^{-35/24} \left({R_\star \over R_\odot}\right) ,
\end{multline} 
meaning that this shocked gas is, typically, photon starved (recall that $\beta/\beta_c< 1$ for their to be a disk upon remnant return). {Note that this result is independent of the change in radius of the stellar remnant, which is an example of a general result that the photon physics in these collisions is independent of the physical size of the returning body \citep[see][]{Mummery25QPE}. } In general, in a purely photon starved state, the observed temperature is roughly given by $kT_{\rm obs} \approx \eta^2 kT_{\rm BB} \approx {\cal O}(100)$ eV, i.e., this collision would produce an X-ray flare. 

As well as photon starvation, however, {other important radiative transfer effects}  can be important in this regime. {Multiple} inverse Compton {scatterings of low energy photons will generally raise the observed temperature of the emission provided both that} the {typical} Compton $y$ parameter exceeds unity $(y \approx (\kappa\Sigma)^2 kT/m_ec^2 \gg 1)$, {and that these low-energy photons are not absorbed by free-free processes before they reach high energies}. {We find that the condition $y \gg 1$ is amply satisfied for the TDE disk parameter range of interest (owing to the very high optical depth $\tau$ of a dense TDE disk), while the second condition is often not satisfied.}  Following \cite{Nakar10}, we compute the  parameter  $y_{\rm max} $using 
\begin{equation}
y_{\rm max} = 3 \left({\rho_{\rm sh} \over 10^{-9} \, {\rm g}\, {\rm cm}^{-3}}\right)^{-1/2} \, \left({k T_{\rm sh} \over 100 \, {\rm eV}}\right)^{9/4} , 
\end{equation}
{noting that free-free absorption plays an important role when $y_{\rm max} \gtrsim 1$}. This expression evaluates to 
\begin{multline}
y_{\rm max} \approx 15 \, {f_d^{1/16} }{\cal A}^{5/16}  \left({\beta\over\beta_c}\right)^{(10+q)/16} \left({h\over r}\right)^{-1/16} \beta_c^{10/16} \\ (1+e_{\rm orb})^{9/16}\left({M_\star \over M_\odot}\right)^{13/48}  \left({R_\star \over R_{\odot}}\right)^{-5/8}\left({M_\bullet \over 10^6 M_\odot}\right)^{17/48}  , 
\end{multline}
meaning that in general {the absorption of photons by free-free processes} is important for the gas shock heated in a remnant-disk collision (a result again independent of the properties of the object on an EMRI). {This effect} limits the peak temperature reached by the photon starved gas through a logarithmic factor $\xi$
\begin{equation}
\xi = {1\over 2} \ln(y_{\rm max}) \left[{8\over 5} + \ln(y_{\rm max})\right] ,
\end{equation}
with for typical parameters is $\xi \gtrsim 3$. The observed temperature, when taking into account both effects discussed above   is approximately given by \citep{Nakar10}
\begin{equation}
k T_{\rm obs} \approx k T_{\rm BB} \times 
\begin{cases}
&1, \quad \quad\quad \,\,\,  \eta < 1, \quad {\rm or} \quad   \xi > \eta > 1, \\
&\eta^2 , \quad\quad \quad  \eta > 1 \quad {\rm and} \quad  \xi < 1,\\
& \left({\eta /  \xi}\right)^{2} , \quad \eta >  \xi > 1.  
 \end{cases}
\end{equation} 
We find for typical parameters that this temperature is roughly $kT_{\rm obs} \sim {\cal O}(100)$ eV, although this result is sensitive to unknowns such as $\beta_c, {\cal A}$, etc. 

In this limit however simple analytical estimates of the resulting character of the emission may well be insufficient, and numerical radiative transfer calculations \citep[i.e.,][]{Vurm24} are of value. We expect detailed simulations to confirm that indeed these flares would be (formally) detectable in X-ray bands. 

\section{Conclusions and future observations}\label{conc}
In this paper we have examined some key differences between the observed properties of the disks formed in full and partial tidal disruption events, focusing in particular on the properties of their X-ray emission. We demonstrate that, assuming tidal disruption event disks behave like scaled-up versions of X-ray binary disks, partial tidal disruption events should undergo X-ray state transitions on a much more rapid evolutionary timescale when compared to full $\beta/\beta_c \geq 1$ disruptions. The scaling, roughly $t_{\rm tr} \sim \beta^{9/4}$, is sufficiently strong that a partial TDE may undergo state transitions in a time as short as $\sim {\cal O}(100)$ days. These results are in good accord with the small number \citep[][]{Wevers21, Liu23, Saxton25} of known sources, who all show this property. This result is therefore supportive of the general paradigm that TDE disks behave just like large black hole mass analogues of X-ray binary disks \citep{Mum21, MumBalb21b}, with their associated spectral state properties. 

{We note here that this paradigm of TDEs as large black hole mass analogues of X-ray binaries suggests a final multi-wavelength observational test, particularly relevant for partial TDEs.  Namely, during the X-ray binary transition to a harder accretion state \citep{Maccarone03,Vahdat2019} radio emission is often detected, interpreted as the switching on of a compact radio bright jet \citep{Fender04}. A natural consequence of the shorter evolutionary time taken for a partial TDE disk to reach the transitional Eddington ratio $f_{\rm edd} \sim 0.01$ would be for partial TDEs to show an enhanced detection rate of radio emission (as a full disruption can comfortably stay above 1\% Eddington for thousands of days, outside of the observational window of typical radio follow up campaigns). Such a statement could be interpreted either as (i) repeating partial TDEs represent particularly promising sources for dedicated radio follow up, or (ii) that radio observations represent possibly the best observational test of this paradigm. A follow up paper shall explore the radio properties of the small handful of repeating partial TDE sources known thus far.   }

We demonstrate that the UV plateau luminosity roughly $\sim$ 1000 days post disruption shows only a weak dependence on $\beta$, with $\beta/\beta_c = 1$ disruptions brightest (as can be understood analytically). At significantly later times ($\sim$ decades) partial disruptions show a more pronounced $\beta$ dependence, although the disks are likely to have undergone a state transitions by this point, which may well modify their behavior. Provided a partial TDE is observed on timescales longer than the decay of the initial flare (typically $\sim$ 1 year), and shorter than the state transitional timescale (also plausibly $\sim 1$ year and therefore certainly not guaranteed), the scaling relationships between black hole mass and UV luminosity should still hold \citep{Mummery_et_al_2024}.  We note that the object AT2018fyk \citep{Wevers21} highlights the possible difficulties here, with the transition to the quiescent state (which took $\sim 600$ days) substantially modifying the UV properties of the disk, although the emission between $t \sim 300-600$ days was in keeping with expectations. 

Finally, we consider the possibility that the returning stellar remnant could produce an observationally relevant interaction with the pre-existing disk formed from its previous passage. We find that the gas shock-heated in these collisions do produce a relatively bright flare, with typical luminosity $L_{\rm flare} \sim 10^{42}$ erg/s, which would likely be detectable in X-ray bands $kT_{\rm obs} \sim 100$ eV. However, these flares are very short lived $t_{\rm flare} \sim 0.1$ hour, significantly shorter than the typical time between stellar passages $t_{\rm repeat} \sim$ months$-$years, and therefore we deem it unlikely they will be detected, except in extremely fortuitous circumstances.

\section*{Acknowledgments}
I would like to thank the referee for detailed comments which improved the manuscript. 
This work was supported by a Leverhulme Trust International Professorship grant [number LIP-202-014]. A.M. acknowledges support from the Ambrose Monell Foundation, the W.M. Keck Foundation and the John N. Bahcall Fellowship Fund at the Institute for Advanced Study.  For the purpose of Open Access, AM has applied a CC BY public copyright license to any Author Accepted Manuscript version arising from this submission. 
 
\section*{Data availability}
The observational data  used in support of this manuscript (Appendix \ref{app}) was taken from the publicly available {\tt manyTDE} data base, which can be accessed at https://github.com/sjoertvv/manyTDE. Numerical scripts will be shared upon request to the corresponding author. 

\bibliographystyle{mnras}
\bibliography{andy}

\appendix

\section{The decay index and luminosity of optical flares in tidal disruption events}\label{app}

{It is well understood that the time evolution of the fallback rate (the rate at which mass stripped from the star returns to orbital pericentre) differs between full and partial disruptions \citep{Coughlin19}. Partial disruptions have mass fallback rates which decay more quickly at late times. {\it If} the late-time decay index of the mass fallback rate set an observationally relevant luminosity then one could attempt to distinguish between partial and full disruptions by fitting a power-law decay to that luminosity    }
\begin{equation}
    L \propto t^p,
\end{equation}
{and associating values $p=-9/4$ with partial disruptions and $p=-5/3$ with full disruptions. }

{Of course, this technique pre-supposes that the evolution of the luminosity one can observe is set only by the rate at which material returns to pericentre, with all of the other physics in the luminosity generation process happening instantaneously.  }

\begin{figure}
    \centering
    \includegraphics[width=0.95\linewidth]{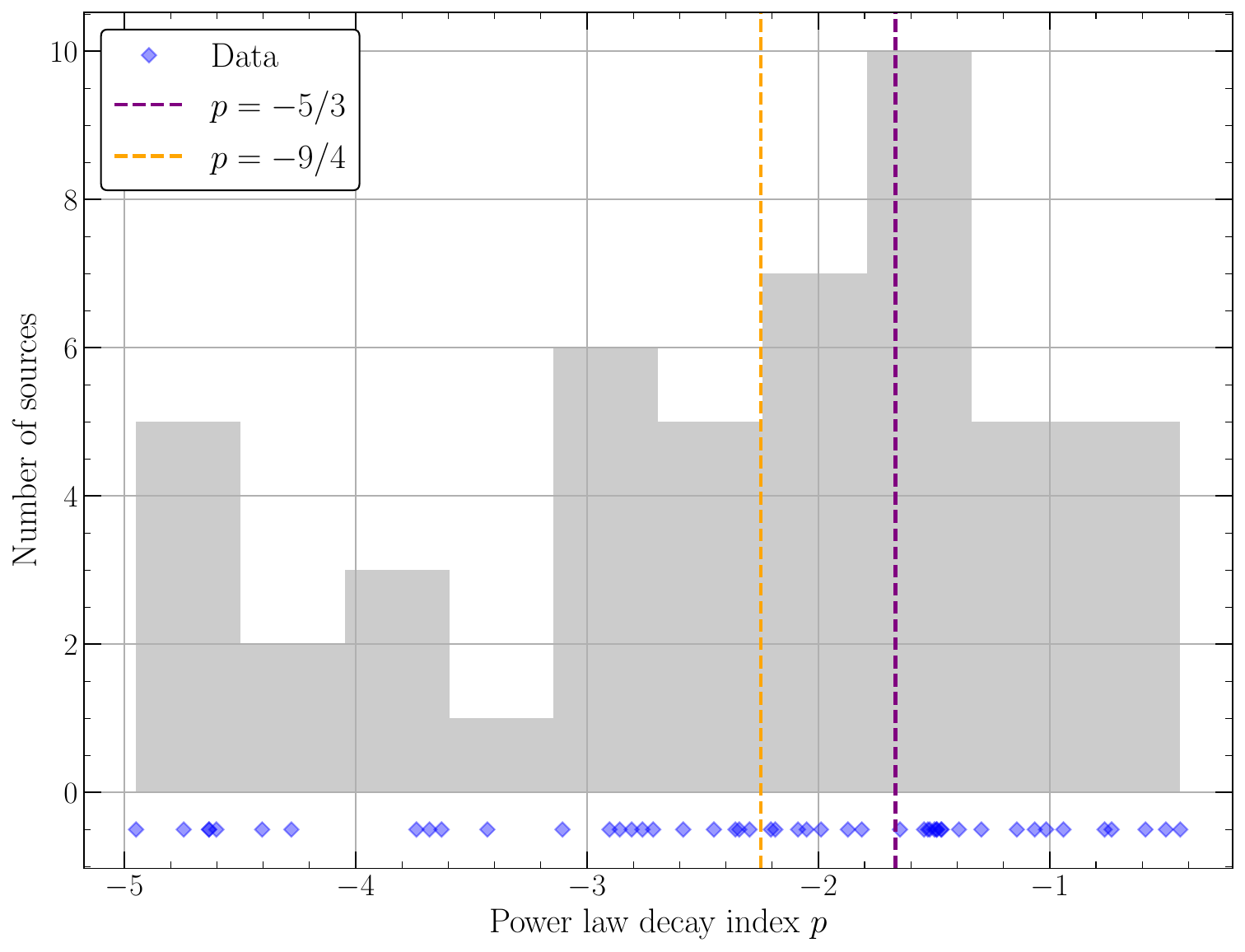}
    \caption{Optical emission decay indices inferred from a large sample of ZTF TDEs (see text for more information). As the distribution is not cleanly bimodal around $-9/4$ and $-5/3$ it is clear that the observed decay index inferred from optical light curves cannot be used to distinguish between different types of stellar disruptions.  }
    \label{fig:index}
\end{figure}

{This supposition can be tested with the large sample of tidal disruption events discovered to date. The light curves of which are publicly available\footnote{https://github.com/sjoertvv/manyTDE}. Fitting a simple powerlaw decay to the post-peak light curves observed across all optical/UV bands \citep[using the standard assumption that the spectrum of the early flare is described by a blackbody function, see][for a full description of the data reduction and model fitting process]{Mummery_et_al_2024} results in an extremely broad distribution of decay indices (Figure \ref{fig:index})\footnote{I am extremely grateful to Sjoert van Velzen for performing these light curve fits and for providing the best-fitting decay indices.}.  }

{As this distribution in decay indices is not cleanly bimodal around $-9/4$ and $-5/3$ it is clear that the observed decay index inferred from optical light curves cannot be used to distinguish between different types of stellar disruptions. Indeed the observed distribution of indices $p$ is, in a statistical sense, perfectly well described by a normal distribution with mean $\mu = -2$ and standard deviation $\sigma = 1$ (with a two sample KS test returning a $p$-value of 0.18, suggesting insufficient evidence to reject the null hypothesis).   }

{We stress that the suggestion in this Appendix is not that the fallback rate does not differ between partial and full disruptions. The suggestion is also not that that the observed distribution of indices $p$ represents any interesting insight into the physics of mass fallback in tidal disruption events. The argument here is that in between mass returning to pericentre and optical/UV emission being produced/observed, additional physics clearly plays a role in setting the observed luminosity evolution, such that the index found from powerlaw fits to optical light curves provides minimal clean information with which to probe the detailed physics of the event.    }

{It is also worth stressing that fallback arguments struggle to reproduce other observed properties of TDE flares, and that this is not limited to the time evolution of the flare itself. {Indeed, fallback arguments typically overestimate the amplitude of the observed emission (often to a significant degree), and the observed scaling of the amplitude of emission with black hole mass differs from that of the fallback rate. } {The following discussion is in effect a brief overview of a much more detailed discussion in \cite{Mummery25cal}, but with the extension to partial TDEs. }

To be explicit, invoking a simple fallback ``luminosity'', as is often done, of $L=\eta \dot M_{\rm fb} c^2$ would imply   }
\begin{multline}\label{nonsense}
    L_{\rm peak} \approx  1 \times 10^{46} \, {\rm erg/s} \,\,\left({M_\bullet \over 10^7M_\odot}\right)^{-1/2} \\  \left({\eta \over 0.1}\right) \left({M_\star \over M_\odot}\right)^2 \left({R_\star \over R_\odot}\right)^{-3/2} ,
\end{multline}
{see, for example, equation 3 of \cite{Wevers23fyk}.} 

{The physics of fallback is of course more complicated than this, particularly for partial TDEs. To perform a more concrete analysis we turn to the numerical simulations of \cite{LawSmith20}, who calibrate the peak fallback rate seen in simulations using realistic stellar structures. They find  }
\begin{equation}
    \dot M_{\rm fb, pk} = 2 f(\beta/\beta_c)\, M_\odot \, {\rm yr}^{-1} \, \left({M_\bullet \over 10^6M_\odot}\right)^{-1/2} \left({M_\star \over M_\odot}\right) ,
\end{equation}
{where we have extended the results of \cite{LawSmith20} to different black hole masses using the canonical $t_{\rm fb} \propto M_\bullet^{1/2}$ result \citep[e.g.,][]{Bandopadhyay+24}. The function $f(\beta/\beta_c)$ is given in the Appendix of \cite{LawSmith20} and varies from $f_{\rm min}\simeq 10^{-2}$, $f_{\rm max}\simeq 2$. We therefore compute the expected fallback luminosity by }
\begin{equation}\label{eq:fb}
    L_{\rm peak} = \eta \dot M_{\rm fb, pk} c^2,
\end{equation}
{where we take the canonical $\eta = 0.1$, and vary over the full range of $\beta/\beta_c$ that was found in \cite{LawSmith20} to result in any disruption at all. }

\begin{figure}
    \centering
    \includegraphics[width=0.95\linewidth]{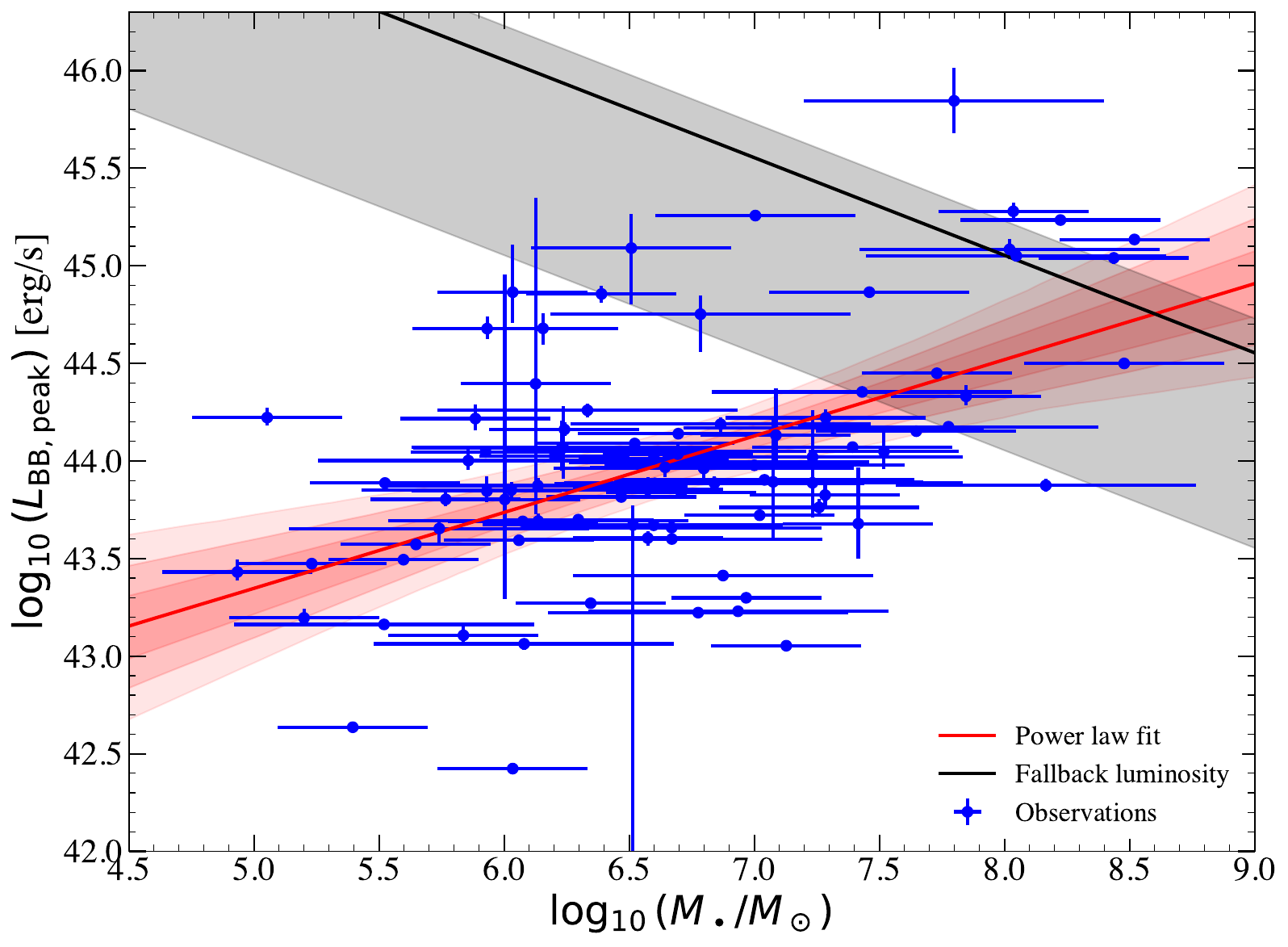}
    \caption{ The peak optical/UV bolometric luminosity inferred from spectral fitting of the ZTF TDE light curves, against the black hole mass inferred from the \citealt{Greene20} scaling relationship and published values of TDE host galaxy velocity dispersions $\sigma$, bulge mass $M_{\rm bulge}$ or galaxy mass $M_{\rm gal}$ (see text). The observed correlation between and the peak bolometric luminosity of the TDE optical/UV flare and the black hole mass in the center of the event is clear to see from visual inspection. The red curve shows the posterior median correlation, while the three shaded regions show $1, 2$ and $3\sigma$ contours respectively.  The peak fallback luminosity for partial and  full TDEs is shown by the black curve and shaded region. }
    \label{fig:index2}
\end{figure}

{The black hole mass scaling and typical amplitude of this relationship are in tension with observations (Figure \ref{fig:index2}).  }

\begin{figure}
    \centering
    \includegraphics[width=0.95\linewidth]{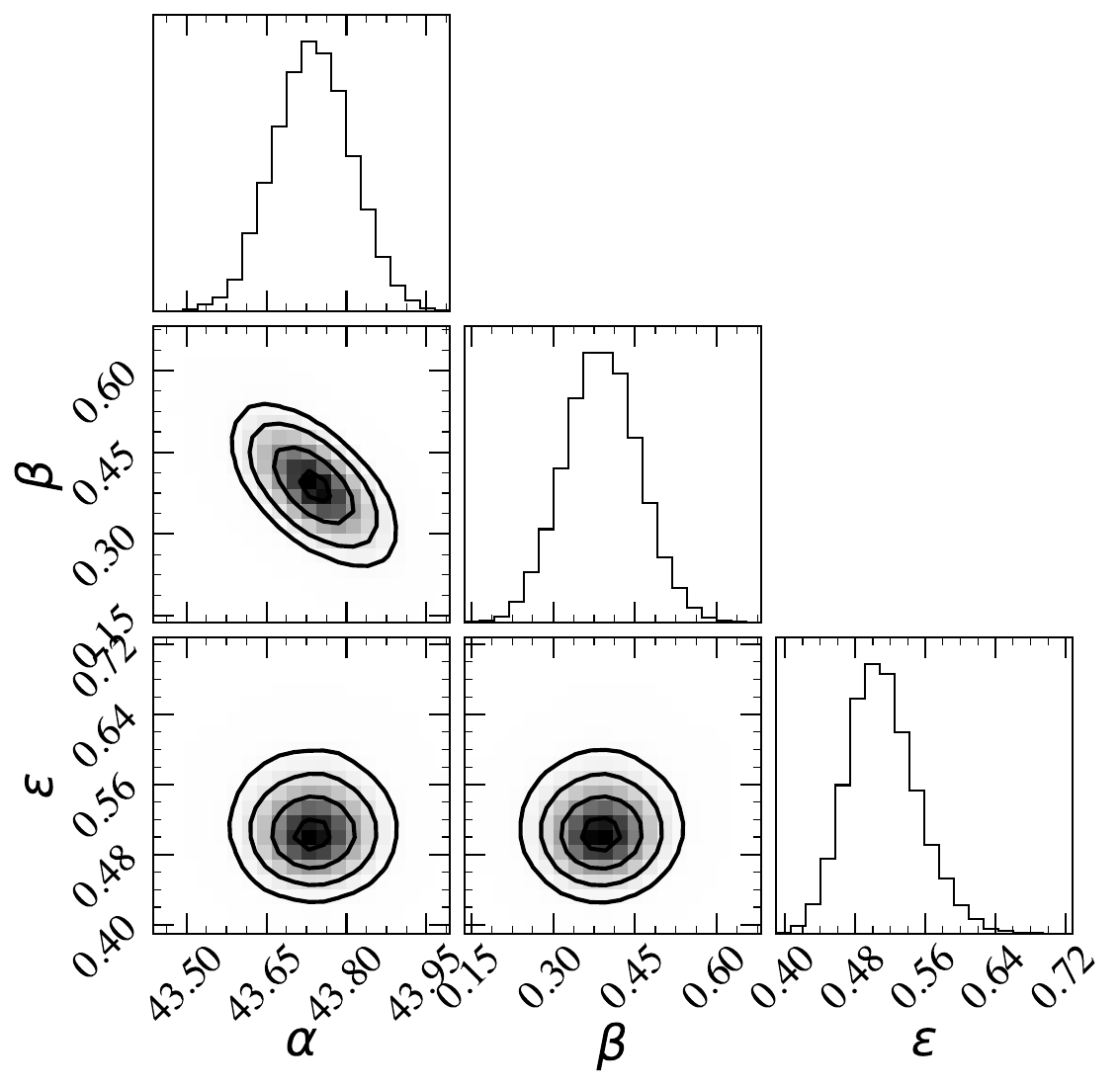}
    \caption{ Corner plot for a power-law fit to the optical/UV peak blackbody luminosity, against black hole mass. A statistically significant positive correlation between $L_{\rm pk}$ and $M_\bullet$ is found. For efficient reprocessing of the fallback rate one would expect $\alpha \simeq 46$, and $\beta = -1/2$.   }
    \label{fig:index3}
\end{figure}

{In Figure \ref{fig:index2} we plot the peak optical/UV bolometric luminosity inferred from spectral fitting of the ZTF TDE light curves, against the black hole mass inferred from the \cite{Greene20} scaling relationship and published values of TDE host galaxy velocity dispersions $\sigma$\footnote{All of these measurements are available in the {\tt manyTDE} database, along with the lightcurves themselves.}, and if $\sigma$ is not available we use the bulge masses presented in \cite{Ramsden25}. If neither the velocity dispersion or bulge mass are available, we use the galaxy mass scaling relationship of \cite{Greene20}.  }

{By a black shaded region we show the predictions of equation \ref{eq:fb}, for a solar mass star spanning the full range of $\beta/\beta_c$ that \cite{LawSmith20} provide. The black curve is the result for a full disruption $\beta \gg \beta_c$.  We plot the data in blue. In red we show the best fitting power law profiles to the data, with the three shaded regions show $1, 2$ and $3\sigma$ contours respectively. }

{To be precise we fit a power-law profile to the data of the following general form }
\begin{equation}\label{power_law_fit}
\log_{10}\left(Y\right) = \alpha + \beta \, \log_{10}\left(X\right) ,
\end{equation}
{where }
\begin{equation}
X \equiv {M_\bullet \over 10^6 M_\odot} ,
\end{equation} 
{and $Y$ is the peak blackbody luminosity in erg/s. The two parameters $\alpha$ and $\beta$ then have simple physical interpretations, namely the (logarithm of the) magnitude of the typical luminosity of a TDE around a $10^6 M_\odot$ black hole, and the power-law index with which this luminosity then scales with black hole mass, respectively.  }

{To understand the intrinsic scatter in these scaling relationships, we incorporate an intrinsic scatter $\epsilon$ into the uncertainty of the luminosity measurements }
\begin{equation}
\left(\delta \log_{10} Y\right)^2 \to \left( \delta \log_{10} Y\right)^2 + \epsilon^2 , 
\end{equation}
{where $\delta \log_{10} Y$ is the uncertainty in the logarithm of each scaling measurement. We then maximize the likelihood }
\begin{multline}
 {\cal L} = - {1\over 2}\sum_{i}  { \left( \log_{10}(Y_i) - \alpha - \beta \log_{10}\left(X_i\right) \right)^2 \over \left(\delta \log_{10}\left(Y_i\right) \right)^2 + \epsilon^2  } \\ + \ln \Big[ 2\pi \left( \left(\delta \log_{10}\left(Y_i \right)  \right)^2 + \epsilon^2 \right)\Big] ,
\end{multline}
{where the summation is over all pairs $(X_i, Y_i)$ of normalised black hole masses and scaling variables. } {Visual inspection of Figure \ref{fig:index2} shows that the luminosity clearly scale positively, and rather strongly, with black hole mass.  }

{Maximising the above likelihood using the {\tt emcee} code \citep{EMCEE}, we infer (Figure \ref{fig:index3}) }
\begin{align}
    &\alpha = 43.7 \pm 0.1, \\
    &\beta = 0.4 \pm 0.1, \label{LpkObs}
\end{align}
{for the peak blackbody luminosity, a $>5\sigma$ discrepancy with the expected index of $-1/2$ from fallback arguments. }

{Of course, one can further suppress the amplitude of the ``fallback luminosity'' (by another order of magnitude) by invoking lower mass stars. However, it seems to the author unlikely that this could be sufficient to suppress the amplitude at $M_\bullet=10^6M_\odot$ to the observed value of $L_{\rm pk} \simeq 10^{43}$ erg/s. }

{Furthermore, this argument would be invoking an inverse Malmquist bias (in effect fine-tuning), whereby flux-limited optical surveys (such as ZTF which provides the bulk of the plotted sample) are only finding the {\it faintest} TDEs, and are missing the population of $L\sim 10^{46}-10^{47}$ erg/s sources (full disruptions around low mass black holes) which should be easy to find. } It is unclear why optical surveys would be missing such objects. 

Model fitting codes, such as {\tt MOSFIT} \citep{Mockler19}, which enforce the fallback rate as a luminosity generating mechanism have a large degree of freedom in fitting, and systematically compensate for this over-prediction of the luminosity. To highlight one example of this parameter compensation explicitly, note that \cite{Nicholl2022} (using a sample of TDEs and {\tt MOSFIT}), recovered an efficiency  scaling of $\eta \propto M_{\bullet}^{0.97 \pm 0.36}$. This can be traced directly to the compensation between the observed luminosity scaling index $\beta = 0.4 \pm 0.1$, and that assumed in {\tt MOSFIT} $\beta = -1/2$, which combined with a tuned efficiency produces $0.47\pm 0.36$, clearly acting solely to compensate for the real scaling in the data.

When a fallback rate code is fit to data therefore, all it is really fitting to is the temporal evolution of the emission (as the luminosity and temperature of the spectrum are effectively free parameters). When one of these codes successfully reproduces the evolution, all this really means is that in a large grid of ``fallback rates'' $\dot M_{\rm fb}(t)$, which can then be stretched freely through $t_{\rm fb} \propto M_\bullet^{1/2}$, there is a profile that looks like the data. It is not clear that this is robust evidence for a fallback powered luminosity mechanism.  It is important to recall that efficient reprocessing of the fallback rate is by no means the only model available for the early optical/UV emission in TDEs, other possibilities include: stream-stream collisions \citep[e.g.,][]{Dai+15,Ryu+20a}, reprocessing of black hole accretion by accretion disk emission by winds/outflows \citep[e.g.,][]{Strubbe&Quataert09,MetzgerStone16},  or the gravitational contraction of an extended quasi-spherical ``envelope'' \citep{Metzger22}, many of which are predicted to evolve with similar {\it temporal} structure, but with different scalings of the peak emission with black hole mass, and which do not expicitly invoke the fallback rate.

\end{document}